\documentclass[conference]{IEEEtran}
\IEEEoverridecommandlockouts
\usepackage{cite}
\usepackage{amsmath,amssymb,amsfonts}
\usepackage{algorithmic}
\usepackage{graphicx}
\usepackage{textcomp}
\usepackage{xcolor}
\def\BibTeX{{\rm B\kern-.05em{\sc i\kern-.025em b}\kern-.08em
    T\kern-.1667em\lower.7ex\hbox{E}\kern-.125emX}}

\usepackage{amsthm, amsmath}
\usepackage{graphicx}
\usepackage{enumitem}
\newtheorem{example}{Example}
\usepackage[linesnumbered,ruled,vlined]{algorithm2e}
\usepackage{balance}  
\newtheorem{proposition}{Proposition}

\usepackage{caption}
\usepackage[skip=1\baselineskip]{caption}
\usepackage{tikz}
 
\usetikzlibrary{calc}
\usetikzlibrary{arrows}
\usepackage{pgfplots}
\usepackage{subfigure}

\newtheorem{lemma}{Lemma}
\usepackage[linesnumbered,ruled]{algorithm2e}
\usepackage{array, booktabs, makecell}
\usepackage{enumitem}
\usepackage[T1]{fontenc}

\usepackage{wrapfig}
\usepackage{lscape}
\usepackage{rotating}
\usepackage{epstopdf}

\setcellgapes{3pt}

\begin{document}

\title{Solving 2-MAXSAT in Polynomial Time: a Proof of $\textit{P}$ = $\textit{NP}$\\
}

\author{Yangjun Chen*

\IEEEcompsocitemizethanks{\IEEEcompsocthanksitem *Y. Chen is with the Department
of Applied Computer Science, The University of Winnpeg, Manitoba, Canada, R3B 2E9.\protect\\
Home page: http://www.acs.uwinnipeg.ca/ychen2
}
\thanks{The article is a modification and extension of a conference paper: Y. Chen, The  2-MAXSAT Problem Can Be Solved in Polynomial Time, in Proc. CSCI2022, IEEE, Dec. 14-16, 2022, Las Vegas, USA, pp. 473-480.

This work is supported by NSERC, Canada, 239074-01 (242523).}
}

\maketitle

\begin{abstract}
By the MAXSAT problem, we are given a set $V$ of $m$ variables and a collection $C$ of $n$ clauses over $V$, i.e., a conjunctive normal form ($\textit{CNF}$) formula. We will seek a truth assignment to maximize the number of satisfied clauses in $C$. This problem is $\textit{NP}$-complete even for its restricted version, the 2-maxsat problem, by which every clause contains at most 2 literals. In this paper, we discuss an efficient algorithm to solve this problem. Its main idea is to transform the 2-maxsat problem into a related problem of maximizing satisfied conjunctions in a disjunctive normal form ($\textit{DNF}$) formula $D$. We then represent all those truth assignments for a conjunction $d$ as a graph (called a $p$*-graph), under each of which $d$ evaluates to $\textit{true}$. In this way, our task becomes finding a maximum number of $p$*-graphs, which have a same $\textit{root-to-leaf}$ path. For this purpose, we organize  all the $p$*-graphs for the conjunctions in $D$ into a trie-like structure. By exploring the structure and recursively its substructures (with each corresponding to a subgraph dynamically built up by integrating some $p$*-subgraphs), the algorithm can find a maximum set of satisfied conjunctions in $D$ in polynomial time. Its worst-case time complexity is bounded by $\mathcal{O}$($n^2(n+m)^4$). This provides in fact a proof of $P$ = $\textit{NP}$.
\end{abstract}


\section{Introduction}
\IEEEPARstart{T}{he} satisfiability problem is perhaps one of the most well-studied problems that arise in many areas of discrete optimization, such as artificial intelligence, mathematical logic, and combinatorial optimization, just to name a few. Given a set $V$ of Boolean ($\textit{true}$/$\textit{false}$) variables and a collection $C$ of clauses over $V$, or say, a logic formula in $\textit{CNF}$ (Conjunctive Normal Form), the satisfiability problem is to determine if there is a truth assignment that satisfies all clauses in $C$  \cite{cook:1971}. The problem is $\textit{NP}$-complete even when every clause in $C$ has at most three literals \cite{Even:1976}. The maximum satisfiability (MAXSAT) problem is an optimization version of satisfiabiltiy that seeks a truth assignment to maximize the number of satisfied clauses \cite{Johnson:1974}. This problem is also $\textit{NP}$-complete even for its restricted version, the so-called 2-MAXSAT problem, by which every clause in $C$ has at most two literals \cite{Garey:1976}. Its application can be seen in an extensive biliography \cite{Djenouri:2017,Garey:1976,Kohli:1994,Kugel:2012,Li:2010,Li:2012,Richard:1974,Shang:2020}.

Over the past several decades, a lot of research on the MAXSAT problem has been conducted. Almost all of them are the approximation methods \cite{Johnson:1974, Krentel:1988, Papadimitriou:1994, Vazirani:1994, Argelich:2013, Dumitrescu:2016}, such as (1-1/$e$)-approximation, 3/4-approximation\cite{Vazirani:1994}, as well as the method based on the integer linear programming \cite{Kemppainen:2020}. The only algorithms for exact solution are discussed in \cite{Zhang:2003, xiao:2022}.  The worst-case time complexity of  \cite{Zhang:2003} is  bounded by $\mathcal{O}$($b$2$^m$), where $b$ is the maximum number of the occurrences of any variable in the clauses of $C$, while the worst-case time complexity of \cite{xiao:2022} is bounded by $\textit{max}$$\{$$\mathcal{O}$($2^m$), $\mathcal{O}$*($1.2989^n$)$\}$. In both algorithms, the traditional branch-and-bound method is used for solving the satisfiability problem, which will search for a solution by letting a variable (or a literal) be 1 or 0. As shown in \cite{Impagliazzo:2001}, any algorithm based on branch-and-bound runs in $\mathcal{O}$*($c^m$) time with $c$ $\geq$ 2.

In this paper, we discuss a polynomial time algorithm to solve the 2-MAXSAT problem. Its worst-case time complexity is bounded by $\mathcal{O}$($n^2(n+m)^4$), where $n$ and $m$ are the numbers of clauses and the number of variables in $C$, respectively. Thus, our algorithm is in fact a proof of $P$ = $\textit{NP}$.

The main idea behind our algorithm can be summarized as follows.

\begin{enumerate}
\item{Given a collection $C$ of $n$ clauses over a set of variables $V$ with each containing at most 2 literals. Construct a formula $D$ over another set of variables $U$, but in $\textit{DNF}$ (Disjunctive Normal Form), containing 2$n$ conjunctions with each of them having at most 2 literals such that there is a truth assignment for $V$ that satisfies  at least $n$* $\leq$ $n$ clauses in $C$ if and only if there is a truth assignment for $U$ that satisfies at least $n$* conjunctions in $D$.
}
\item{For each $D_i$ in $D$ ($i$ $\in$ $\{$1, ..., 2$n$$\}$), construct a graph, called a $p$*-graph to represent all those truth assignments $\sigma$ of variables such that under $\sigma$ $D_i$ evaluates to $\textit{true}$.
}
\item{Organize the $p$*-graphs for all $D_i$'s into a trie-like graph $G$. Searching $G$ bottom up recursively, we can find a maximum subset of satisfied conjunctions in polynomial time.
}
\end{enumerate}

The organization of the rest of this paper is as follow. First, in Section II, we restate the definition of the 2-MAXSAT problem and show how to reduce it to a problem that seeks a truth assignment to maximize the number of satisfied conjunctions in a formula in $\textit{DNF}$. Then, we discuss a basic algorithm and its improvement in Section III. Section IV is devoted to the analysis of  the time complexity. Finally, a short conclusion is set forth in Section V.


\section{2-MAXSAT Problem}\label{section2}

We will deal solely with Boolean variables (that is, those which are either $\textit{true}$ or $\textit{false}$), which we will denote by $c_1$, $c_2$, etc. A literal is defined as either
a variable or the negation of a variable (e.g., $c_7$, $\neg$$c_{11}$ are literals). A literal $\neg$$c_i$ is $\textit{true}$ if the variable $c_i$ is $\textit{false}$. A clause is
defined as the OR of some literals, written as  ($l_1$ $\lor$ $l_2$ $\lor$ .... $\lor$ $l_k$) for some $k$ ($\geq$ 1), where each $l_i$ (1 $\leq$ $i$ $\leq$ $k$) is a literal, as illustrated in  $\neg$$c_1$ $\lor$ $c_{11}$. We say
that a Boolean formula is in conjunctive normal form ($\textit{CNF}$) if it is presented
as an AND of clauses: $C_1$ $\land$ ... $\land$ $C_n$ ($n$ $\geq$ 1). For example, ($\neg$$c_1$ $\lor$ $c_7$ $\lor$ $\neg$$c_{11}$) $\land$ ($c_5$ $\lor$ $\neg$$c_2$ $\lor$ $\neg$$c_3$) is in $\textit{CNF}$. In addition, a disjunctive normal form ($\textit{DNF}$) is an OR of conjunctions: $D_1$ $\lor$ $D_2$ $\lor$ ... $\lor$ $D_m$ ($m$ $\geq$ 1). For instance, ($c_1$ $\land$ $c_2$) $\lor$ ($\neg$$c_1$ $\land$ $c_{11}$) is in $\textit{DNF}$.

Finally, the MAXSAT problem is to find an assignment to the variables of
a Boolean formula in $\textit{CNF}$ such that the maximum number of clauses are set to $\textit{true}$, or are satisfied. Formally:

\vspace*{0.2cm}
2-MAXSAT
\begin{itemize}
\item{Instance: A finite set $V$ of variables, a Boolean formula $C$ = $C_1$ $\land$ ... $\land$ $C_n$ in $\textit{CNF}$ over $V$ such that each $C_i$ has 0 < $|C_i|$ $\leq$ 2 literals ($i$ = 1, ..., $n$), and a positive integer $n$* $\leq$ $n$.}
\item{Question: Is there a truth assignment for $V$ that satisfies at least $n$* clauses?}
\end{itemize}




In terms of \cite{Garey:1976}, the 2-MAXSAT problem is $\textit{NP}$-complete.

To find a truth assignment $\sigma$ such that the number of clauses set to $true$ is maximized under $\sigma$, we can try all the possible assignments, and count the satisfied clauses as discussed in \cite{Li:2012}, by which bounds are set up to cut short branches. We may also use a heuristic method to find an approximate solution to the problem as described in \cite{Johnson:1974}.

In this paper, we propose a quite different method, by which for $C$ = $C_1$ $\land$ ... $\land$ $C_n$, we will consider another formula $D$ in $\textit{DNF}$ constructed as follows.

Let $C_i$ = $c_{i1}$ $\lor$ $c_{i2}$ be a clause in $C$, where $c_{i1}$ and $c_{i2}$ denote either variables in $V$ or their negations. For $C_i$, define a variable $x_i$. and a pair of conjunctions: $D_{i1}$, $D_{i2}$, where

\vspace*{0.2cm}

\hspace*{1cm} $D_{i1}$ = $c_{i1}$ $\land$ $x_i$,

\hspace*{1cm} $D_{i2}$ = $c_{i2}$ $\land$ $\neg$$x_i$.

\vspace*{0.2cm}

Let $D$ = $D_{11}$ $\lor$ $D_{12}$ $\lor$ $D_{21}$ $\lor$ $D_{22}$ $\lor$ ... $\lor$ $D_{n1}$ $\lor$ $D_{n2}$. Then, given an instance of the 2-MAXSAT problem defined over a variable set $V$ and a collection $C$ of $n$ clauses, we can construct a logic formula $D$ in $\textit{DNF}$ over the set $V$ $\cup$ $X$ in polynomial time, where  $X$ = $\{$$x_1$, ..., $x_n$$\}$. $D$ has $m$ = 2$n$ conjunctions.

Concerning the relationship of $C$ and $D$, we have the following proposition.

\begin{proposition}
Let $C$ and $D$ be a formula in $\textit{CNF}$ and a formula in $\textit{DNF}$ defined above, respectively. No less than $n$* clauses in $C$ can be satisfied by a truth assignment for $V$ if and only if no less than $n$* conjunctions in $D$ can be satisfied by some truth assignment for $V$ $\cup$ $X$.
\label{proposition1}
\end{proposition}

\begin{proof}
Consider every pair of conjunctions in $D$: $D_{i1}$ = $c_{i1}$ $\land$ $x_i$ and $D_{i2}$ = $c_{i2}$ $\land$ $\neg$$x_i$ ($i$ $\in$ $\{$1, ..., $n$$\}$). Clearly, under any truth assignment for the variables in $V$ $\cup$ $X$, at most one of $D_{i1}$ and $D_{i2}$ can be satisfied. If $x_i$ = $\textit{true}$, we have $D_{i1}$ = $c_{i1}$ and $D_{i2}$ = $\textit{false}$. If $x_i$ = $\textit{false}$, we have $D_{i2}$ = $c_{i2}$ and $D_{i1}$ = $\textit{false}$.

"$\Rightarrow$" Suppose there exists a truth assignment $\sigma$ for $C$ that satisfies $p$ $\geq$ $n$* clauses in $C$. Without loss of generality, assume that the $p$ clauses are $C_{1}$, $C_{2}$, ..., $C_{p}$.

Then, similar to Theorem 1 of \cite{Kohli:1994}, we can find a truth assignment $\tilde{\sigma}$ for $D$, satisfying the following condition:

For each $C_{j}$ = $c_{j1}$ $\lor$ $c_{j2}$ ($j$ = 1, ..., $p$), if $c_{j1}$ is $\textit{true}$ and $c_{j2}$ is $\textit{false}$ under $\sigma$, (1) set both $c_{j1}$ and $x_j$ to $\textit{true}$ for $\tilde{\sigma}$. If $c_{j1}$ is $\textit{false}$ and $c_{j2}$ is $\textit{true}$ under $\sigma$, (2) set $c_{j2}$ to $\textit{true}$, but $x_j$ to $\textit{false}$ for $\tilde{\sigma}$. If both  $c_{j1}$ and $c_{j2}$ are $\textit{true}$, do (1) or (2) arbitrarily.

Obviously, we have at least $n$* conjunctions in $D$ satisfied under $\tilde{\sigma}$.

"$\Leftarrow$" We now suppose that a truth assignment $\tilde{\sigma}$ for $D$ with  $q$ $\geq$ $n$* conjunctions in $D$ satisfied. Again, assume that those $q$ conjunctions are $D_{1b_1}$, $D_{2b_2}$, ..., $D_{qb_q}$, where each $b_j$ ($j$ = 1, ..., $q$) is 1 or 2. 

Then, we can find a truth assignment $\sigma$ for $C$, satisfying the following condition:

For each $D_{jb_j}$ ($j$ = 1, ..., $q$), if $b_{j}$ = 1, set $c_{j1}$ to $\textit{true}$ for $\sigma$; if $b_{j}$ = 2, set $c_{j2}$ to $\textit{true}$ for $\sigma$.

Clearly, under $\sigma$, we have at lease $n$* clauses in $C$ satisfied.

The above discussion shows that the proposition holds.
\end{proof}

Proposition 1 demonstrates that the 2-MAXSAT problem can be transformed, in polynomial time, to a problem to find a maximum number of conjunctions in a logic formula in $\textit{DNF}$.

As an example, consider the following logic formula in $\textit{CNF}$:

\begin{equation}
\begin{array}{ll}
	C =  C_1 \land C_2 \land C_3\\
\hspace*{0.385cm} =  (c_1 \lor c_2) \land (c_2 \lor \neg c_3) \land (c_3 \lor \neg c_1). 
\end{array} 
\end{equation}

Under the truth assignment $\sigma$ = $\{$$c_1$ = 1, $c_2$ = 1, $c_3$ = 1$\}$, $C$ evaluates to $\textit{true}$, i.e., $C_i$ = 1 for $i$ = 1, 2, 3. Thus, $n$* = 3.

For $C$, we will generate another formula $D$, but in $\textit{DNF}$, according to the above discussion:

\begin{equation}
\begin{array}{ll}
D = D_{11} \lor D_{12} \lor D_{21} \lor D_{22} \lor D_{31} \lor D_{32}\\
 \hspace*{0.39cm}=  (c_1 \land c_4) \lor (c_2 \land \neg c_4) \lor \\
\hspace*{0.77cm}  (c_2 \land c_5) \lor (\neg c_3 \land \neg c_5) \lor \\
\hspace*{0.77cm}  (c_3 \land c_6) \lor (\neg c_1 \land \neg c_6).
\end{array} 
\end{equation}

$D$ contains 6 conjunctions and 6 variables. According to Proposition~\ref{proposition1}, $D$ should also have at least $n$* = 3 conjunctions which evaluates to $\textit{true}$ under some truth assignment. In the opposite, if $D$ has at least 3 satisfied conjunctions under a truth assignment, then $C$ should have at least three clauses satisfied by some truth assignment, too. In fact, it can be seen that under the truth assignment $\tilde{\sigma}$ = $\{$$c_1$ = 1, $c_2$ = 1, $c_3$ = 1, $c_4$ = 1, $c_5$ = 1, $c_6$ = 1$\}$, $D$ has three satisfied conjunctions: $D_{11}$, $D_{21}$, and $D_{31}$, from which the three satisfied clauses in $C$ can be immediately determined.

In the following, we will discuss a polynomial time algorithm to find a maximum set of satisfied conjunctions in any logic formula in $\textit{DNF}$, not only restricted to the case that each conjunction contains up to 2 conjuncts.

\section{Algorithm description}\label{section3}

In this section, we discuss our algorithm. First, we present the main idea in Section~\ref{subsection3.1}. Then, in Section~\ref{subsection3.2}, a recursive algorithm for solving the problem is described in great detail. Next, in Section ~\ref{subsection3.3}, we discuss how this algorithm can be further improved. The running time of the algorithm will be analyzed in the  next section.

\subsection{Main idea}\label{subsection3.1}

To develop an efficient algorithm to find a truth assignment that maximizes the number of satisfied conjunctions in formula $D$ = $D_1$ $\lor$ ..., $\lor$ $D_n$, where each $D_i$ ($i$ = 1, ..., $n$) is a conjunction of variables $c$ ($\in$ $V$),  we need to represent each $D_i$ as a  sequence of variables (referred to as a variable sequence). For this purpose, we introduce a new notation:
\vspace*{0.2cm}

 ($c_j$, *) = $c_j$ $\lor$ $\neg$$c_j$ = $\textit{true}$,
\vspace*{0.2cm}

\noindent which will be inserted into $D_i$ to represent any missing variable $c_j$ $\notin$ $D_i$ (i.e., $c_j$ $\in$ $V$, but not appearing in $D_i$). Obviously, the truth value of each $D_i$ remains unchanged.

In this way, the above $D$ can be rewritten as a new formula in $\textit{DNF}$ as follows:

\begin{equation}
\begin{array}{ll}
D =  D_{1} \lor D_{2} \lor D_{3} \lor D_{4} \lor D_{5} \lor D_{6}\\
\hspace*{0.4cm} = (c_1 \land (c_2, *) \land (c_3, *) \land c_4 \land (c_5, *) \land (c_6, *)) \lor \\
    \hspace*{0.755cm}  ((c_1, *) \land c_2\land  (c_3, *) \land \neg c_4 \land (c_5, *) \land (c_6, *))  \lor \\
     \hspace*{0.755cm} ((c_1, *) \land c_2 \land (c_3, *) \land (c_4, *) \land c_5 \land (c_6, *)) \lor&\\
\hspace*{0.755cm} ((c_1, *) \land (c_2, *) \land \neg c_3 \land (c_4, *) \land \neg c_5 \land (c_6, *)) \lor\\
\hspace*{0.755cm} ((c_1, *) \land (c_2, *) \land c_3 \land (c_4, *) \land (c_5, *) \land c_6) \lor\\
\hspace*{0.755cm} (\neg c_1 \land (c_2, *) \land (c_3, *) \land (c_4, *)  \land (c_5, *) \land \neg c_6).
\end{array}
\end{equation}

Doing this enables us  to represent each $D_i$ as a variable sequence, but with all the negative literals being removed. It is because if the variable in a negative literal is set to $\textit{true}$, the corresponding conjunction  $D_i$ must be $\textit{false}$, and our goal is to establish a graph in which each node represents a variable and each ($\textit{root-to-leaf}$) path $p$ corresponds to a truth assignment satisfying  $D_i$ (by which any variable on $p$ is set $\textit{true}$ while all those varibles not on $p$ are set $\textit{false}$). Obviousely, in such a graph, any variable appearing in a negative literal should not be involved since any path through such a variable corresponds to a truth assignment not satisfying $D_i$.

\vspace{0.2cm}
See Table~\ref{table1} for illustration.

\begin{table*}[h]
\centering
\caption{Conjunctions represented as sorted variable sequences.}
\vspace{-0.3cm}
\resizebox{14cm}{!}{
\begin{tabular}{|l|l|l|} \hline
 conjunction&variable sequences&sorted variable sequences\\ \hline \hline  
 $D_1$&$c_1$.($c_2$, *).($c_3$, *).$c_4$.$(c_5, *)$.($c_6$, *)&$\#.$($c_2$, *).($c_3$, *).$c_1$.$c_4$.$(c_5, *)$.($c_6$, *).$\$$\\
$D_2$&($c_1$, *).$c_2$.($c_3$, *).($c_5$, *).($c_6$, *) &$\#.$$c_2$.($c_3$, *).($c_1$, *).($c_5$, *).($c_6$, *).$\$$ \\
$D_3$&($c_1$, *).$c_2$.($c_3$, *).($c_4$, *).$c_5$.($c_6$, *) &$\#.$$c_2$.($c_3$, *).($c_1$, *).($c_4$, *).$c_5$.($c_6$, *).$\$$\\
$D_4$&($c_1$, *).($c_2$, *).($c_4$, *).($c_6$, *) &$\#.$($c_2$, *).($c_1$, *).($c_4$, *).($c_6$, *).$\$$\\
$D_5$&($c_1$, *).($c_2$, *).$c_3$.($c_4$, *).($c_5$, *).$c_6$&$\#.$($c_2$, *).$c_3$.($c_1$, *).($c_4$, *).($c_5$, *).$c_6$.$\$$\\
$D_6$&($c_2$, *).($c_3$, *).($c_4$, *).($c_5$, *) &$\#.$($c_2$, *).($c_3$, *).($c_4$, *).($c_5$, *).$\$$\\
\hline\end{tabular}
}
\label{table1}\\
\large
\end{table*}

First, we pay attention to the variable sequence for $D_2$ (the second sequence in the second column of Table ~\ref{table1}), in which the negative literal $\neg$$c_4$ (in $D_2$) is elimilated. In the same way, you can check all the other variable sequences.

Now it is easy for us to compute the appearance frequencies of different variables in the variable sequences, by which each ($c$, *) is counted as a single appearance of $c$ while any negative literals are not considered, as illustrated in Table~\ref{table2}, in which we show the appearance frequencies of all the variables in the above $D$.

\begin{table}[ht]
\centering
\caption{Appearance frequencies of variables.}
\vspace{-0.3cm}
\resizebox{8.5cm}{!}{
\begin{tabular}{|l|l|l|l|l|l|l|} \hline
 variables&$c_1$&$c_2$&$c_3$&$c_4$&$c_5$&$c_6$\\ \hline \hline 
 appearance frequencies&5/6&6/6&5/6&5/6&5/6&5/6\\
\hline\end{tabular}
}
\label{table2}\\
\large
\end{table}

According to the variable appearance frequencies, we will impose a global ordering over all variables in $D$ such that the most frequent variables appear first, but with ties broken arbitrarily. For instance, for the $D$ shown above, we can specify a global ordering like this: $c_2$ $\to$ $c_3$ $\to$ $c_1$ $\to$ $c_4$ $\to$ $c_5$ $\to$ $c_6$. Here, $c_2$ is most frequent and then appears first. The other variables have the same frequency. So, we simply impose a fixed order on them: $c_3$ $\to$ $c_1$ $\to$ $c_4$ $\to$ $c_5$ $\to$ $c_6$.

Following this general ordering, each conjunction $D_i$ in $D$ can be represented as a sorted variable sequence as illustrated in the third column of Table~\ref{table1}, where the variables in a sequence are ordered in terms of their appearance frequencies such that more frequent variables appear before less frequent ones. In addition, a  start symbol $\#$ and an end symbol $\$$ are used as $\textit{sentinals}$ for technical convenience. In fact, any global ordering of variables works well (i.e., you can specify any global ordering of variables), based on which a graph representation of assignments can be established. However, ordering variables according to their appearance frequencies can greatly improve the efficiency when 
searching a graph constructed over all the variable sequences for conjunctions in $D$ to find solusions since more variables from different conjunctions can be merged together. 

Later on, by a variable sequence, we always mean a sorted variable sequence. Also, we will use $D_i$  and the variable sequence for $D_i$ interchangeably without causing any confusion.

In addition, for our algorithm, we need to introduce a graph structure to represent all the truth assignments for each $D_i$ ($i$ = 1, ..., $n$) (called a $p$*-graph), under each of which $D_i$ evaluates to $\textit{true}$. In the following, however, we first define a simple concept of $p$-graphs for ease of explanation.

\vspace{0.2cm}
\noindent $\textbf{Definition}$ $\textbf{1.}$ ($p$-graph) Let $\alpha$ $=$ $d_0$$d_1$ ... $d_k$$d_{k+1}$ be a variable sequence representing a $D_i$ in $D$ as described above (with $d_0$ $=$ $\#$, $d_{k+1}$ $=$ $\$$, and each $d_i$ with $i$ $\in$ $\{$1, ..., $k$$\}$ is a variable or a a pair of the form ($c$, *), where $c$ is a variable). A $p$-graph over $\alpha$ is a directed graph, in which there is a node for each $d_j$ ($j$ $=$ $0$, ..., $k$ $+$ $1$); and an  edge for ($d_j$, $d_{j+1}$) for each $j$ $\in$ $\{$$0$, $1$, ..., $k$$\}$. In addition, for each $d_i$ with $i$ $\in$ $\{$$1$, ..., $k$$\}$, if it is a pair of the form  ($c$, *), an extra edge connecting $d_{j-1}$ to $d_{j+1}$ is added.

\begin{figure*}[ht]
\centering
\includegraphics[width=150mm,scale=0.6]{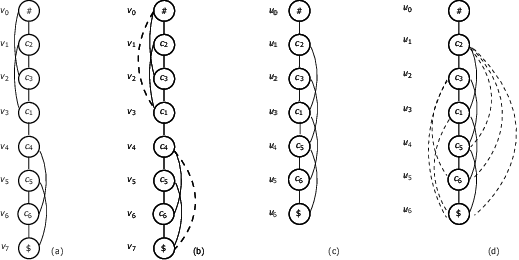}
\caption{Illustration for $p$-graphs and $p$*-graphs.}
\label{fig1}
\end{figure*}

\vspace{0.2cm}
In Fig.~\ref{fig1}(a), we show such a $p$-graph for $D_1$ = $d_0$$d_1$$d_2$$d_3$$d_4$$d_5$$d_6$$d_7$ = $\#.$($c_2$, *).($c_3$, *).$c_1$.$c_4$.($c_5$, *).($c_6$, *).$\$$. Beside a main path going through all the  variables in $D_1$, there are four off-path edges (edges not on the main path), referred to as $\textit{spans}$ attached to the main path, corresponding to ($c_2$, *), ($c_3$, *),  ($c_5$, *), and ($c_6$, *), respectively. Each span is represented by the subpath covered by it. For example, we will use the subpath $<$$v_0$, $v_1$, $v_2$$>$ (subpath going three nodes: $v_0$, $v_1$, $v_2$) to stand for the span connecting $v_0$ and $v_2$; $<$$v_1$, $v_2$, $v_3$$>$ for the span connecting $v_1$ and $v_3$; $<$$v_4$, $v_5$, $v_6$$>$ for the span connecting $v_4$ and $v_6$, and  $<$$v_5$, $v_6$, $v_7$$>$ for the span connecting $v_6$ and $v_7$. By using spans, the meaning of `*'s (it is either 0 or 1) is appropriately represented since along a span we can bypass the corresponding variable (then its value is set to 0) while along an edge on the main path we go through the corresponding variable (then its value is set to 1).

In fact, what we want is to represent,  in an efficient way, all those truth assignments for each $D_i$ ($i$ = 1, ..., $n$), under each of which $D_i$ evaluates to $\textit{true}$. However, $p$-graphs fail to do so since when we go through from a node $v$ to another node $u$ through a span, $u$ must be selected. If $u$ represents a ($c$, *) for some variable name $c$, the meaning of this `*' is not properly rendered.  It is because ($c$, *) indicates that $c$ is optional, but going through a span from $v$ to ($c$, *) makes $c$ always selected. So, the notation ($c$, *), which is used to indicate that $c$ is optional, is not correctly implemented.

For this reason, we introduce another concept, $p$*-graphs, described as below.

Let $s_1$ = $<$$v_1$, ..., $v_k$$>$ and  $s_2$ = $<$$u_1$, ..., $u_l$$>$ be two spans attached onto a same main path. We say, $s_1$ and $s_2$ are overlapped, if $u_1$ = $v_j$ for some $j$ $\in$ $\{$1, ..., $k$ - 1$\}$, or if $v_1$ = $u_{j'}$ for some $j'$ $\in$ $\{$1, ..., $l$ - 1$\}$. For example, in Fig.~\ref{fig1}(a), $<$$v_0$, $v_1$, $v_2$$>$ and $<$$v_1$, $v_2$, $v_3$$>$ are overlapped. $<$$v_4$ $v_5$, $v_6$$>$ and $<$$v_5$, $v_6$, $v_7$$>$ are also overlapped. 

Here, we notice that if we had one more span, $<$$v_3$, $v_4$, $v_5$$>$, for example, it would be connected to $<$$v_1$, $v_2$, $v_3$$>$, but not overlapped with $<$$v_1$, $v_2$, $v_3$$>$. Being aware of this difference is important since the overlapped spans imply the consecutive `*'s, just like $<$$v_0$, $v_1$, $v_2$$>$ and $<$$v_1$, $v_2$, $v_3$$>$, which correspond to two consecutive `*'s: ($c_2$, *) and ($c_3$, *). Therefore, the overlapped spans exhibit some kind of $\textit{transitivity}$. That is, if $s_1$ and $s_2$ are two overlapped spans, then $s_1$ $\cup$ $s_2$ must be a new, but bigger span. Applying this operation to all the spans over a $p$-path, we will get a '$\textit{transitive closure}$' of overlapped spans. 

Let $S$ be the set of all spans over the main path $p$ for a certain conjunction. The transitive closure of $S$, denoted as $S$*, is another set of spans $S$* = $\{$$s_1$, $s_2$, ..., $s_l$$\}$ for some $l$, which contains the whole $S$ and is with each $s_i$ satisfying one of the following two conditions:

1. $s_i$ $\in$ $S$, or

2. There exist $j$, $k$ ($\neq$ $i$) such that $s_j$ and $s_k$ are overlapped and $s_i$ = $s_j$ $\cup$ $s_k$.

Based on the above discussion, we give the following definition.

\noindent $\textbf{Definition}$ $\textbf{2.}$ ($p$*-graph) Let $P$ be a $p$-graph. Let $p$ be its main path and $S$ be the set of all spans over $p$. Denote by $S$* the `transitive closure' of $S$. Then, the $p$*-graph with respect to $P$ is the union of $p$ and $S$*, denoted as $P$* $=$ $p$ $\cup$ $S$*.

In Fig.~\ref{fig1}(b), we show the $p$*-graph with respect to the $p$-graph given in Fig.~\ref{fig1}(a). 

As another example, consider $D_2$ =  $\#.$$c_2$.($c_3$, *).($c_1$, *).($c_5$, *).($c_6$, *).$\$$. Its $p$-graph is shown in  Fig.~\ref{fig1}(c) and its $p$*-graph in Fig.~\ref{fig1}(d), in which we notice that we have span $<$$u_2$, $u_3$, $u_4$, $u_5$$>$ (representing two consecutive `*'s) due to two overlapped spans: $<$$u_2$, $u_3$, $u_4$$>$ and $<$$u_3$, $u_4$, $u_5$$>$. Further, we have span $<$$u_2$, $u_3$, $u_4$, $u_5$, $u_6$$>$ (representing three consecutive `*'s) due to  $<$$u_2$, $u_3$, $u_4$, $u_5$$>$ and  $<$$u_4$, $u_5$, $u_6$$>$. In the same way, we can check all the other spans in Fig.~\ref{fig1}(d).

The purpose of the $p$*-graph for a certain conjunction $D_i$ is to represent all the truth assignments, under each of which $D_i$ evaluates to $\textit{true}$. Specifically, in $P$* each $\textit{root}$-to-$\textit{leaf}$ path $p$ corresponds to a truth assignment, by which each variable on $p$ is set to $\textit{true}$ while any other variables (not on $p$) are set $\textit{false}$.

Concerning $p$*-graphs, we have the following lemma.

\begin{lemma}
Let $P$* be a $p$*-graph for a conjunction $D_i$ (represented as a variable sequence) in $D$. Then, any path from $\#$ to $\$$ in $P$* represents a truth assignment, under which $D_i$ evaluate to $\textit{true}$.
\label{lemma1}
\end{lemma}

\begin{proof}
(1) Corresponding to any truth assignment $\sigma$, under which $D_i$ evaluates to $true$, there is definitely a path from $\#$ to $\$$ in $p$*-graph. First, we note that under such a truth assignment each variable in a positive literal must be set to 1 and definitely appear on the main path. Any optional variable (represented by ($c$, *) for some variable $c$), if it takes 1, will appear on the main path; otherwise, will go along a span bypassing the corresponding node on the main path. Especially, we may have more than one consecutive `*'s that are set 0 under the truth assignment. They can be represented by a span that is the union of the corresponding overlapped spans. In addition, any variable in a negative literal must take value 0 and will not appear on any path. Therefore, for $\sigma$ we must have a path representing it.

(2) Each path from $\#$ to $\$$ represents a truth assignment, under which $D_i$ evaluates to $\textit{true}$. To see this, we observe that each path consists of several edges on the main path and several spans. Especially, any such path must go through every variable in a positive literal since for each of them there is no span covering it. But each span stands for a `*' or more than one consecutive `*', corresponding to a single optional or several consecutive optional variables. Since any variable in a negative literal does not appear on the path, it must have value $\textit{false}$. So, under the truth assignment represented by the path $D_i$ evaluates to $\textit{true}$.
\end{proof}

For example, in Fig.~\ref{fig1}(b), the path: $v_0$ $\to$ $v_3$ $\to$ $v_4$ $\to$ $v_5$ $\to$ $v_7$ represents a truth assignment: $\{$$c_1$ = 1, $c_2$ = 0, $c_3$ = 0, $c_4$ = 1, $c_5$ = 1, $c_6$ = 0$\}$, under which $D_1$ evaluates to $\textit{true}$. In Fig. ~\ref{fig1}(d), the path: $u_0$ $\to$ $u_1$ $\to$ $u_2$ $\to$ $u_6$ represents another truth assignment: $\{$$c_1$ = 0, $c_2$ = 1, $c_3$ = 1, $c_4$ = 0, $c_5$ = 0, $c_6$ = 0$\}$, under which $D_2$ evaluates to $\textit{true}$. We can examine all the paths in these two graphs and find that Lemma 1 always  holds for them.

From the above discussion, we can see that our main task now is to find a maximum number of $p$*-graphs (each representing a concjunction in a $\textit{DNF}$ formula) such that each of them contains a same $\textit{root-to-leaf}$ path.

\subsection{Algorithm}\label{subsection3.2}

To find a truth assignment to maximize the number of satisfied conjunctions in $D$, we will first construct a $\textit{trie-like}$ structure $G$ over $D$, and then search $G$ bottom-up recursively to find answers.

Let $P_1$*, $P_2$*, ..., $P_n$* be all the $p$*-graphs constructed for all conjunctions in $D$. Let $p_j$ and $S_j$* ($j$ = 1, ..., $n$) be the main path of $P_j$* and the transitive closure over its spans, respectively. We will construct $G$ in two steps.

In the first step, we will establish a $\textit{trie}$ \cite{CM:1993, Kruth:1975}, denoted as $T$ = $trie$($R$) over $R$ = $\{$$p_1$, ...,  $p_n$$\}$ as  follows. 

If $|R|$ = 0, $trie$($R$) is, of course, empty. For $|R|$ = 1, $trie$($R$) is a single node. If $|R|$ $>$ 1, $R$ is split into  $r$ (possibly empty) subsets $R_1$, $R_2$, …, $R_r$ so that each $R_i$ ($i$ = 1, …, $r$) contains all those sequences with the same first variable name. The tries: $trie(R_1$), $trie(R_2$), …, $trie(R_r$) are constructed in the same way except that at the $k$th step, the splitting of sets is based on the $k$th variable name (along the global ordering of variables). They are then connected from their respective roots to a single node to create $trie$($R$).

\begin{figure}[ht]
\centering
\includegraphics[width=85mm,scale=0.8]{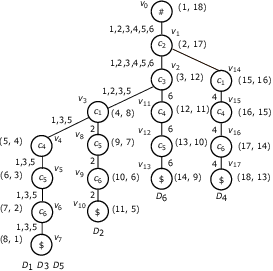} 
\caption{A trie and tree encoding.}
\label{fig5-3}
\end{figure}

In Fig.~\ref{fig5-3}, we show the trie constructed for the variable sequences given in the third column of Table~\ref{table1}. In such a trie, special attention should be paid to all the leaf nodes each labeled with $\$$, representing a conjunction (or a subset of conjunctions), which can be satisfied under the truth assignment represented by the corresponding main path. For example, the  subset $\{$$D_1$, $D_3$, $D_5$$\}$ associated with $v_7$ (a leaf node) is satisfiable under the truth assignment represented by the path from $v_0$ to $v_7$. Such a path is also called a tree path. 

The main advantage of tries is to cluster common parts of variable sequences together to avoid possible repeated checking. Then, if variable sequences are sorted according to their appearance frequencies, more variables will be clustered and more time can be saved. More importantly, this idea can also be applied to the variable subsequences (as will be seen later), over which some dynamical tries can be recursively constructed, leading to a polynomial-time algorithm for solving the problem. It is because any dynamically constructed sub-trie will enable us to reduce a possible exponential-time checking to a single path search.

Each node $v$ in the trie stands for a variable $c$, referred to as the label of $v$ and denodeted as $l$($v$) = $c$; and each edge $e$ is referred to as a tree edge, labeled with a set of integers representing all the variable sequences going through $e$, denoted as $s(e)$. For example, $s(v_0, v_1)$ = $\{$1, 2, 3, 4, 5, 6$\}$. It is because all the variable sequences given in Table~\ref{table1} need to pass through this edge to reach their respective leaf nodes. In the same way, you can check all the other labels associated with tree edges.

Concerning the tree paths, we have the following lemma.

\begin{lemma}
Let $T$ be a trie created over the $p$*-graphs' main paths for all the variable sequences in $D$. Let $p$ = $v_0$ $\xrightarrow{s_1}$ $v_1$ ...  $\xrightarrow{s_k}$ $v_k$ be a root-to-leaf path in $T$. Let $A$ be the subset of conjunctions associated with $v_k$. Then, we have $A$ = $s_1$ $\cap$ ... $\cap$ $s_k$, which
is a subset of conjunctions satisfiable by the truth assignment represented by $p$.
\label{lemma2}
\end{lemma}

Finally, we will associate each node $v$ in the trie $T$ with a pair of numbers ($\textit{pre}$, $\textit{post}$) to speed up recognizing ancestor/descendant relationships of nodes in $T$, where $\textit{pre}$ is the order number of $v$ when searching $T$ in preorder and $\textit{post}$ is the order number of $v$ when searching $T$ in postorder.

These two numbers can be used to characterize the ancestor/descendant relationships in $T$ as follows.
\vspace{0.2cm}
\begin{itemize}
\item[-]{ Let $v$ and $v’$ be two nodes in $T$. Then, $v’$ is a descendant of $v$ iff $\textit{pre}$($v’$) $>$ $\textit{pre}$($v$) and $\textit{post}$($v’$) $<$ $\textit{post}$($v$).
}
\end{itemize}
\vspace{0.2cm}

For the proof of this property of any tree, see Exercise 2.3.2-20 in \cite{Kruth:1969}.

For instance, by checking the label associated with $v_2$ against the label for $v_9$ in Fig.~\ref{fig5-3}, we see that $v_2$ is an ancestor of $v_9$ in terms of this property. Specifically, $v_2$’s label is (3, 12) and $v_9$’s label is (10, 6), and we have 3 < 10 and 12 > 6. We also see that since the pairs associated with $v_{14}$ and $v_6$ do not satisfy  the property, $v_{14}$ must not be an ancestor of $v_{6}$ and $\textit{vice versa}$. In addition, for technical convenience, we can regard a node as a descendant of the node itself. Thus, the condition for the ancestor/descendant relationship can be slightly changed as follows:

\vspace{0.2cm}
\begin{itemize}
\item[-]{ Let $v$ and $v’$ be two nodes in $T$. Then, $v’$ is a descendant of $v$ iff $\textit{pre}$($v’$) $\geq$ $\textit{pre}$($v$) and $\textit{post}$($v’$) $\leq$ $\textit{post}$($v$).
}
\end{itemize}

\vspace{0.2cm}



\begin{figure}[ht]
\centering
\includegraphics[width=75mm,scale=0.65]{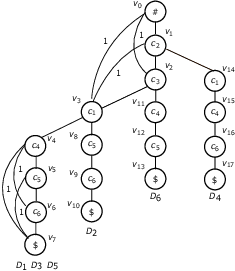} 
\caption{Illustration for adding spans to a trie.}
\label{fig5-4}
\end{figure}
\begin{figure*}[ht]
\centering
\includegraphics[width=120mm,scale=0.8]{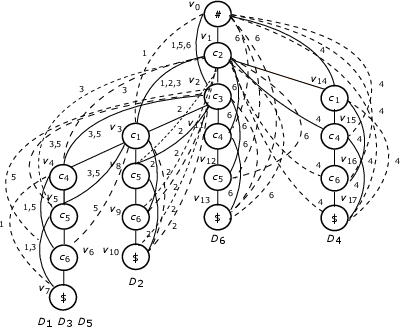} 
\caption{A trie-like graph $G$.}
\label{fig3}
\end{figure*}

In the second step, we will add all $S_i$* ($i$ = 1, ..., $n$) to the trie $T$ to construct a trie-like graph $G$, as illustrated in Fig.~\ref{fig3}. (In Fig.~\ref{fig5-4}, we show the graph after all the spans in the $p$*-graph for $D_1$ are added.) This trie-like graph is constructed for all the variable sequences given in Table~\ref{table1}, in which each span is also associated with a set of numbers used to indicate what variable sequences the span belongs to. For example, the span $<$$v_0$, $v_1$, $v_2$$>$ (in Fig.~\ref{fig3})  is associated with three numbers: 1, 5, 6, indicating that the span belongs to 3 conjunctions: $D_1$, $D_5$, and $D_6$. In Fig. ~\ref{fig3}, however, the labels for all tree edges are not shown for a clear illustration.

Also, each $p$*-graph itself is considered to be a simple trie-like graph.

Concerning the paths in a trie-like graph, we have another lemma similar to Lemma~\ref{lemma2}.

\begin{lemma}
Let $G$ be a trie-like graph created over all the variable sequences in $D$. Let $p$ = $v_0$ $\xrightarrow{s_1}$ $v_1$ ...  $\xrightarrow{s_k}$ $v_k$ be a root-to-leaf path in $G$, where some edges can be spans. Let $A$ be the subset of conjunctions associated with $v_k$. Then, $R$ = $s_1$ $\cap$ ... $\cap$ $s_k$ $\cap$ $A$
is satisfiable by the truth assignment represented by $p$.
\label{lemma3}
\end{lemma}

From Fig.~\ref{fig3}, we can see that although the number of truth assignments for $D$ is exponential, they can be represented by a graph with polynomial numbers of nodes and edges. In fact, in a single $p$*-graph over $m$ variables, the number of edges is bounded by $\mathcal{O}$($m^2$). Thus, a trie-like graph over $n$ $p$*-graphs has at most O($nm^2$) edges.

In a next step, we will search $G$ bottom-up (in postorder) to seek all the possible largest subsets of conjunctions which can be satisfied by a certain truth assignment.

First of all, we call each node in $T$ with more than one child a $\textit{branching}$ node. For instance, node $v_3$ with two children $v_4$ and $v_8$ in $G$ shown in Fig.~\ref{fig3} is a branching node. For the same reason,  $v_2$ and $v_1$ are another two branching nodes. But $v_0$ is not a branching node since it has only one child in $T$ (although it has more than one child in $G$.)

Around branching nodes, we have an interesting concept defined below, which is very important for the dynamic construction of trie-like subgraphs and the recursive bottom-up searching of $G$, especially, for finding truth assignments for all those conjunctions, which correspond to paths going through some spans.

\vspace{0.2cm}

\noindent $\textbf{Definition}$ $\textbf{3.}$ (reachable subsets through spans) Let $v$ be a branching node. Let $u$ be a node on the tree path (in $T$) from $\textit{root}$ to $v$ (not including $v$ itself). A reachable subset of $u$ through spans are all those nodes with a same label $c$ in different subgraphs in $G$[$v$] (subgraph of $G$ rooted at $v$) and reachable from $u$ through a span, denoted as $\textit{RS}^{v,u}_s$[$c$], where $s$ is a set containing all the labels associated with the corresponding spans.

\vspace{0.2cm}

For any node $w$ in $\textit{RS}^{v,u}_s$[$c$], node $u$ is also called its $\textit{anchor}$ node while $w$ is called a $\textit{reachable}$ node of $u$. $s$ and $c$ are referred to as the span label and node label of the $\textit{RS}$, respectively.

For instance, for node $v_2$ in Fig. ~\ref{fig3}, which is  on the tree path from $\textit{root}$ to $v_3$ (a branching node), we have two $\textit{RS}$s with respect to $v_3$:

\vspace{0.2cm}
\begin{itemize}
\item[-]{$\textit{RS}^{v_3,v_2}_{\{2,3,5\}}$[$c_5$] = $\{$$v_5$, $v_8$$\}$,
}
\vspace{0.2cm}
\item[-]{$\textit{RS}^{v_3,v_2}_{\{2,5\}}$[$c_6$] = $\{$$v_6$, $v_9$$\}$.
}
\end{itemize}

We have $\textit{RS}^{v_3,v_2}_{\{2,3,5\}}$[$c_5$] due to two spans $v_2$ $\xrightarrow{3, 5}$ $v_5$ and $v_2$ $\xrightarrow{2}$ $v_8$ going out of $v_2$, respectively reaching $v_5$ and $v_8$ in two different $p$*-subgraphs in $G$[$v_3$] with $l$($v_5$) =  $l$($v_8$) = `$c_5$'. We have $\textit{RS}^{v_3,v_2}_{\{2,5\}}$[$c_6$] due to another two spans going out of $v_2$: $v_2$ $\xrightarrow{5}$ $v_6$ and $v_2$ $\xrightarrow{2}$ $v_9$ with $l$($v_6$) = $l$($v_9$) = `$c_6$'. 

By merging the subgraphs rooted at the nodes in an $\textit{RS}$, we can contruct a new trie-like subgraph, and find some more satisfiable conjunctions by searching it recursively.

However, if an $\textit{RS}$ contains only a single node, no new trie-like subgraph can be produced. It is useless and therefore should not be further considered. So, in the subsequent discussion, by an  $\textit{RS}$, we always mean an $\textit{RS}$ with |$\textit{RS}$| $\geq$ 2.

In addition, the definition of this concept for an anchor node $u$ which is a branching node itself is a little bit different from any other node on the tree path (from $\textit{root}$ to $u$). Specifically, each of its $\textit{RS}$s is defined to be a subset of nodes reachable from $u$ via a span or  $\textit{via a tree edge}$. So, for $v_3$ (as an anchor node with respect to  $v_3$ as a branching node) in Fig. ~\ref{fig3} we have:

\vspace{0.2cm}

\begin{itemize}
\item[-]{$\textit{RS}^{v_3,v_3}_{\{2,3,5\}}$[$c_5$] = $\{$$v_5$, $v_8$$\}$,
}
\vspace{0.2cm}
\item[-]{$\textit{RS}^{v_3,v_3}_{\{2,5\}}$[$c_6$] = $\{$$v_6$, $v_9$$\}$.
}
\end{itemize}


\noindent The first one is due to span $v_3$ $\xrightarrow{3,5}$ $v_5$ and tree edge $v_3$ $\xrightarrow{2}$ $v_8$ going out of $v_3$ with  $l$($v_6$) =  $l$($v_8$) = `$c_5$'; and the second is due to two spans
$v_3$ $\xrightarrow{5}$ $v_6$ and $v_3$ $\xrightarrow{2}$ $v_9$ going out of $v_3$ with  $l$($v_6$) =  $l$($v_8$) = `$c_6$'. Here, we notice that the label for the tree edge $v_3$ $\rightarrow$ $v_8$ is 2 since this tree edge belongs to $D_2$ alone (see Fig.~\ref{fig5-3}.)

What we want is to construct a trie-like subgraph, in terms of each $\textit{RS}$,  by merging the subgraphs rooted respectively at the nodes in it. Then, we can try to find some more satisfiable conjunctions in this new subgraph. Obviously, this can be done in a recursive way.

In the following, we will discuss some examples to illustrate the above working process in great detail.

\begin{figure*}[h]
\centering
\includegraphics[width=120mm,scale=0.8]{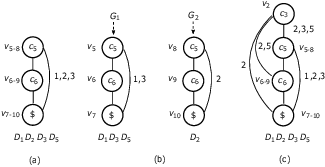}
\caption{Illustration for construction of trie-like subgraphs with respect to $v_3$.}
\label{fig3-2}
\end{figure*}

\begin{example}
In terms of the first $\textit{RS}$ with respect to $v_3$: $\textit{RS}^{v_3,v_2}_{\{2,3,5\}}$[$c_5$] = $\{$$v_5$, $v_8$$\}$, we can construct a trie-like subgraph as shown in Fig. ~\ref{fig3-2}(a) by merging two subgraphs in $G$[$v_3$]: $G_1$ - rooted at $v_5$, and $G_2$ - rooted at $v_8$ (see Fig. ~\ref{fig3-2}(b) for illustration). In this new trie-like subgraph, $v_{5-8}$ is the merging of $v_5$ and $v_8$, $v_{6-9}$ is the merging of $v_6$ and $v_9$, and $v_{7-10}$ is the merging of $v_7$ and $v_{10}$. 
\label{example1}
\end{example}

In this new trie-like subgraph, for technical convenience, a virtual root is added. It is in fact the corresponging anchor node $v_2$ which is connected to the root of the new trie-like subgraph through a virtual edge, whose label is set to be the union of all the corresponding span labels: $\{$2, 3, 5$\}$ = $\{$3, 5$\}$ $\cup$ $\{$2$\}$. (See Fig. ~\ref{fig3-2}(c)). In addition, all the edges from $v_2$ to some other nodes in this trie-like subgraph will be inherited from $G$. (See $v_2$ $\xrightarrow{2,5}$ $v_{6-9}$ and $v_2$ $\xrightarrow{2}$ $v_{7-10}$ in Fig. ~\ref{fig3-2}(c) for illustration).

We also notice that the new merged trie-like subgraph is just a single path (with some spans) and the unique leaf node is associated with $\{$$D_1, D_2, D_3, D_5$$\}$. But $D_1$ will be filtered out by the label associated with the virtual edge $\{$2, 3, 5$\}$.  In this way, what we find is a set of conjunctions $\{$$D_2, D_3, D_5$$\}$ which can be satisfied by a truth assignment represented by a path shown below:


\vspace{0.2cm}

 $v_0$ $\rightarrow$ $v_1$ $\rightarrow$ $v_2$ $\xrightarrow{2,3,5}$ $v_{5-8}$ $\rightarrow$ $v_{6-9}$ $\rightarrow$ $v_{7-10}$,\\

\noindent which consists of two tree paths:\\

\noindent path in $G$:  $v_0$ $\rightarrow$ $v_1$, and 

\noindent path in the graph shown in Fig. ~\ref{fig3-2}(c): $v_{5-8}$ $\rightarrow$ $v_{6-9}$ $\rightarrow$ $v_{7-10}$,\\

\noindent connected by a $\textit{combined}$ span $v_2$ $\xrightarrow{2,3,5}$ $v_{5-8}$, which is the merging of two spans: $v_2$ $\xrightarrow{3,5}$ $v_{5}$, and $v_2$ $\xrightarrow{2}$ $v_{8}$.

\vspace{0.2cm}

The truth assignment represented by this path is $\{$$c_1$ = 0, $c_2$ = 1, $c_3$ = 1, $c_4$ = 0, $c_5$ = 1, $c_6$ = 1$\}$. Here, we notice that $l(v_0)$ = $\#$, $l(v_1)$ = $c_2$, $l(v_2)$ = $c_3$, $l(v_{5-8})$ = $c_5$, $l(v_{6-9})$ = $c_6$, and $l(v_{7-10})$ = $\$$; and any variable on the path has value equal to $\textit{true}$ while any variable not on the path has vaue $\textit{false}$.

Since this trie-like subgraph contains no branching node, no further recursive call will be conducted.

Next, in terms of the second $\textit{RS}$ with respect to $v_3$: {$\textit{RS}^{v_3,v_2}_{\{2,5\}}$[$c_6$] = $\{$$v_6$, $v_9$$\}$, we will create another trie-like subgraph as shown in Fig. ~\ref{fig3-3}(a). This is the merging of two trie-like subgraphs shown in Fig. ~\ref{fig3-3}(b), reachable from $v_2$ through two spans labeled respectively with 2 and 5. Therefore, the label for the virtual edge is $\{$2, 5$\}$. (See Fig. ~\ref{fig3-3}(c)). The leaf node is associated with $\{$$D_1, D_2, D_3, D_5$$\}$. But the result is $\{$$D_2, D_5$$\}$. Both $D_1$ and $D_3$ are filtered out by the label attached  to this virtual edge.

\begin{figure*}[h]
\centering
\includegraphics[width=110mm,scale=0.8]{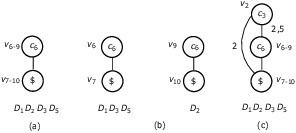}
\caption{Illustration for construction of trie-like subgraphs with respect to $v_3$.}
\label{fig3-3}
\end{figure*}

Again, the corresponding truth assignment is represented by a tree path from $\textit{root}$ to $v_2$ in $G$ plus a path from $v_2$ (the virtual node in the trie-like subgraph shown in Fig. ~\ref{fig3-3}(c)) to $v_{7-10}$:\\

 $v_0$ $\rightarrow$ $v_1$ $\rightarrow$ $v_2$ $\xrightarrow{2,5}$ $v_{6-9}$ $\rightarrow$ $v_{7-10}$,\\

\noindent representing a truth assignment: $\{$$c_1$ = 0, $c_2$ = 1, $c_3$ = 1, $c_4$ = 0, $c_5$ = 0, $c_6$ = 1$\}$ .\\

Finally, for the remaining two $\textit{RS}$s with respect to $v_3$: $\textit{RS}^{v_3,v_3}_{\{2,5\}}$[$c_5$]  = $\{$$v_5$, $v_8$$\}$ and  $\textit{RS}^{v_3,v_3}_{\{2,5\}}$[$c_6$] = $\{$$v_6$, $v_9$$\}$, the same computation as above will be performed, creating the same trie-like subgraphs as those shown in Fig. ~\ref{fig3-2}(a) and Fig. ~\ref{fig3-3}(a).

In the above examples, no recursive graph searching is involved. But in general, since a newly created trie-like subgraph may contain some branching nodes, the recursive graph exploration is normally required. 

\vspace{0.2cm}

The following example helps for illustration.

\vspace{0.2cm}

\begin{example}
If we continually search $G$ bottom-up, the next encountered branching node is $v_2$ (with label $c_3$), with respect to which we have seven $\textit{RS}$s:
\label{example2}
\end{example}

- $\textit{RS}^{v_2,v_1}_{\{3,6\}}$[$c_4$] = $\{$$v_4$, $v_{11}$$\}$,\\

- $\textit{RS}^{v_2,v_1}_{\{2,3,6\}}$[$c_5$] = $\{$$v_5$, $v_8$, $v_{12}$$\}$,\\

- $\textit{RS}^{v_2,v_1}_{\{2,6\}}$[$\$$] = $\{$$v_{10}$, $v_{13}$$\}$,\\

- $\textit{RS}^{v_2,v_2}_{\{3,5,6\}}$[$c_4$] = $\{$$v_4$, $v_{11}$$\}$,\\

- $\textit{RS}^{v_2,v_2}_{\{2,3,5,6\}}$[$c_5$] = $\{$$v_5$, $v_8$, $v_{12}$$\}$,\\

- $\textit{RS}^{v_2,v_2}_{\{2,5\}}$[$c_6$] = $\{$$v_6$, $v_{9}$$\}$,\\

- $\textit{RS}^{v_2,v_2}_{\{2,6\}}$[$\$$] = $\{$$v_{10}$, $v_{13}$$\}$,\\

The first three $\textit{RS}$s are due to the following three sets of spans emitting from $v_1$, respectively  (see Fig. ~\ref{fig3}):\\

 - $\{$$v_1$ $\xrightarrow{3}$ $v_4$, $v_1$ $\xrightarrow{6}$ $v_{11}$$\}$,\\

 - $\{$$v_1$ $\xrightarrow{2}$ $v_8$, $v_1$ $\xrightarrow{3}$ $v_5$, $v_1$ $\xrightarrow{6}$ $v_{12}$$\}$,\\

- $\{$$v_1$ $\xrightarrow{2}$ $v_{10}$, $v_1$ $\xrightarrow{6}$ $v_{13}$$\}$.\\

The rest four $\textit{RS}$s are respectively due to four groups of spans emitting from $v_2$ shown below (note that the anchor $v_2$ is the branching node $v_2$ itself; see Fig. ~\ref{fig3}):\\

- $\{$$v_2$ $\xrightarrow{3,5}$ $v_4$, $v_2$ $\xrightarrow{6}$ $v_{11}$$\}$,\\

 - $\{$$v_2$ $\xrightarrow{3,5}$ $v_5$, $v_2$ $\xrightarrow{2}$ $v_{8}$, $v_2$ $\xrightarrow{6}$ $v_{12}$$\}$,\\

 - $\{$$v_2$ $\xrightarrow{5}$ $v_6$, $v_2$ $\xrightarrow{2}$ $v_{9}$$\}$,\\

 - $\{$$v_2$ $\xrightarrow{2}$ $v_{10}$, $v_2$ $\xrightarrow{6}$ $v_{13}$$\}$.\\

In terms of the first $\textit{RS}$ with respect to $v_2$: $\textit{RS}^{v_2,v_1}_{\{3,6\}}$[$c_4$] = $\{$$v_4$, $v_{11}$$\}$, we will generate a trie-like subgraph as shown in Fig. ~\ref{fig3-4}(a) by merging two subgraphs respectively rooted at $v_4$ and $v_{11}$ in $G$[$v_2$] (see Fig.  ~\ref{fig3-4}(b)). Again, a virtual root ($v_1$) is added as shown in Fig. ~\ref{fig3-4}(c), which is the anchor of $v_4$ and $v_{11}$. The label for the virtual edge is $\{$3, 6$\}$. Therefore, $D_1$ and $D_5$ will be filtered out from the subset $\{$$D_1, D_3, D_5$$\}$ associated with the leaf node $v_7$. This shows that $\{$$D_3$$\}$ can be satisfied by a truth assignment:

\vspace{0.2cm}

$\{$$c_1$ = 0, $c_2$ = 1, $c_3$ = 0, $c_4$ = 1, $c_5$ = 1, $c_6$ = 1$\}$, \\

\noindent represented by a path shown as follows:\\

$v_0$ $\rightarrow$ $v_1$ $\xrightarrow{3,6}$ $v_{4-11}$ $\rightarrow$ $v_{5-12}$ $\rightarrow$ $v_{6}$ $\rightarrow$ $v_{7}$.\\

\begin{figure*}[ht]
\centering
\includegraphics[width=160mm,scale=1]{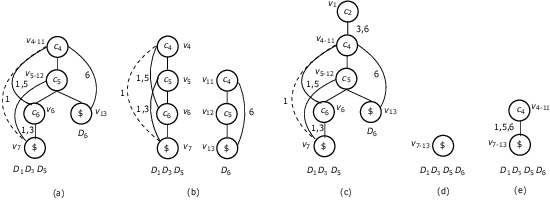}
\caption{Illustration for construction of trie-like subgraphs with respect to $v_2$.}
\label{fig3-4}
\end{figure*}

The subset associated with another leaf node ($v_{13}$) of the subgraph shown in Fig. ~\ref{fig3-4}(c) is $\{$$D_6$$\}$ which can be satisfied by $\{$$c_1$ = 0, $c_2$ = 1, $c_3$ = 0, $c_4$ = 1, $c_5$ = 1, $c_6$ = 0$\}$. The corresponding path is:\\

$v_0$ $\rightarrow$ $v_1$ $\xrightarrow{3,6}$ $v_{4-11}$ $\rightarrow$ $v_{5-12}$ $\rightarrow$ $v_{13}$.\\

However, both the subsets are smaller than $\{$$D_1, D_3, D_5$$\}$ that has already been found before. So, such results needn't be kept around.

In addition, this trie-like subgraph contains a branching node $v_{5-12}$. A further recursive call of the algorithm will then be conducted. For this, we will first construct the corresponding $\textit{RS}$s:\\

- $\textit{RS}^{v_{5-12},v_{5-12}}_{\{1,5,6\}}$[$\$$] = $\{$$v_7$, $v_{13}$$\}$.\\

- $\textit{RS}^{v_{5-12},v_{4-11}}_{\{1,6\}}$[$\$$] = $\{$$v_7$, $v_{13}$$\}$.\\

In terms of the first $\textit{RS}$, we can establish a next trie-like subgraph shown in Fig. ~\ref{fig3-4}(d). It contains only a single node. Adding the corresponding virtual node and virtual edge, we get a subgraph as shown in Fig. ~\ref{fig3-4}(e), from which we can get a satisfiable subset of conjunctions $\{$$D_6$$\}$. Although the subset associated with the unique leaf node of the trie-like subgraph shown in Fig. ~\ref{fig3-4}(e) is $\{$$D_1, D_3, D_5, D_6$$\}$, it is reduced to $\{$$D_6$$\}$ by two labels associated respectively with the two virtual edges on the corresponding path: $\{$3, 6$\}$ $\cup$ $\{$1, 5, 6$\}$ = $\{$6$\}$. (In Fig. ~\ref{fig3-4}(c) and (e), we can observe respectively these two virtual edges: ($v_1$, $v_{4-11}$), and ($v_{4-11}$, $v_{7-13}$)).\\

We notice that this path is:\\

$v_0$ $\rightarrow$ $v_1$ $\xrightarrow{3,6}$ $v_{4-11}$ $\rightarrow$ $v_{5-12}$ $\xrightarrow{1,5,6}$ $v_{7-13}$,\\


\noindent representing the following truth assignment:\\

$\{$$c_1$ = 0, $c_2$ = 1, $c_3$ = 0, $c_4$ = 1, $c_5$ = 1, $c_6$ = 0$\}$.\\

In the same way, we can handle the second $\textit{RS}$ with respect to $v_{5-12}$.

For the second $\textit{RS}$ with respect to $v_2$: $\textit{RS}^{v_2,v_1}_{\{2,3,6\}}$[$c_5$] = $\{$$v_5$, $v_8$, $v_{12}$$\}$, a trie-like subgraph as shown in Fig. ~\ref{fig3-5} will be build up, which also contains a branching node $v_{5-8-12}$, possibly leading to some more recursive calls.

For all the remaining five $\textit{RS}$s with respect to $v_2$: $\textit{RS}^{v_2,v_1}_{\{2,6\}}$[$\$$] = $\{$$v_{10}$, $v_{13}$$\}$, $\textit{RS}^{v_2,v_2}_{\{3,5,6\}}$[$c_4$] = $\{$$v_4$, $v_{11}$$\}$, $\textit{RS}^{v_2,v_2}_{\{2,3,5,6\}}$[$c_5$] = $\{$$v_5$, $v_8$, $v_{12}$$\}$, $\textit{RS}^{v_2,v_2}_{\{2,5\}}$[$c_6$] = $\{$$v_6$, $v_{9}$$\}$, $\textit{RS}^{v_2,v_2}_{\{2,6\}}$[$\$$] = $\{$$v_{10}$, $v_{13}$$\}$, the computation is similar to the above 2 examples.\\

\begin{figure}[ht]
\centering
\includegraphics[width=35mm,scale=1]{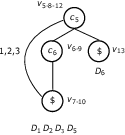}
\caption{An trie-like subgraph.}
\label{fig3-5}
\end{figure}

Now, we study a third example to show a kind of redundancy, by which identical recursive calls may appear along different series of recursive calls. 

\begin{example}
In this example, we consider the third branching node $v_1$ in $G$. With respect to this branching node, we have three $\textit{RS}$s for $v_0$:
\label{example3}
\end{example}

- $\textit{RS}^{v_1,v_0}_{\{1,4\}}$[$c_1$] = $\{$$v_{3}$, $v_{14}$$\}$,\\

- $\textit{RS}^{v_1,v_0}_{\{4,6\}}$[$c_4$] = $\{$$v_{11}$, $v_{15}$$\}$,\\

- $\textit{RS}^{v_1,v_0}_{\{4,6\}}$[$\$$] = $\{$$v_{13}$, $v_{17}$$\}$,\\

We have them due to the following three span groups emitting from $v_0$:\\

 - $\{$$v_0$ $\xrightarrow{1}$ $v_3$, $v_0$ $\xrightarrow{4}$ $v_{14}$$\}$,\\

 - $\{$$v_0$ $\xrightarrow{6}$ $v_{11}$, $v_0$ $\xrightarrow{4}$ $v_{15}$$\}$,\\

 - $\{$$v_0$ $\xrightarrow{6}$ $v_{13}$, $v_0$ $\xrightarrow{4}$ $v_{17}$$\}$,\\

According to the three $\textit{RS}$s, we will create three trie-like graphs as shown in Fig. ~\ref{fig3-7}(a), (b), and (c), respectively.

\begin{figure*}[ht]
\centering
\includegraphics[width=150mm,scale=1]{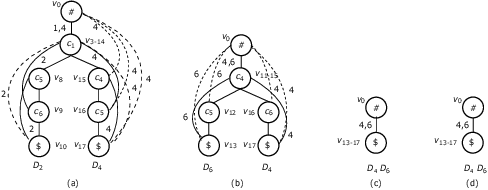}
\caption{Illustration for construction of trie-like subgraphs with respect to $v_1$.}
\label{fig3-7}
\end{figure*}

In Fig. ~\ref{fig3-7}(a), $D_2$ associated with the leaf node $v_{10}$ cannot be satisfied by the corresponding path since it cannot pass through the label for the virtual edge ($v_0$, $v_{3-14}$): $\{$1, 4$\}$. But  $D_4$ associated with another leaf node $v_{17}$ is able to penetrate the label and satisfied by the corresponding path:\\

$v_0$ $\xrightarrow{1,4}$ $v_{3-14}$ $\rightarrow$ $v_{15}$ $\rightarrow$ $v_{16}$ $\rightarrow$ $v_{17}$,\\

\noindent representing a truth assignment:\\

$\{$$c_1$ = 1, $c_2$ = 0, $c_3$ = 0, $c_4$ = 1, $c_5$ = 1, $c_6$ = 0$\}$.\\

Besides, the subgraph shown in Fig. ~\ref{fig3-7}(a) also contains a branching node $v_{3-14}$. With repect to it, we have three $\textit{RS}$s for $v_0$: $\textit{RS}^{v_1,v_0}_{\{4\}}$[$c_4$] = $\{$$v_{15}$$\}$, $\textit{RS}^{v_1,v_0}_{\{4\}}$[$c_6$] = $\{$$v_{16}$$\}$, and $\textit{RS}^{v_1,v_0}_{\{4\}}$[$\$$] = $\{$$v_{17}$$\}$. However, all of them are of size equal to 1 and will be simply ignored.

Next, from the subgraph shown in Fig. ~\ref{fig3-7}(b), we can get another two satisfiable subsets $\{$$D_6$$\}$ and $\{$$D_4$$\}$, associated respectively with its two leaf nodes. $D_6$ can be satisfied by the path:

\vspace{0.2cm}

$v_0$ $\xrightarrow{4,6}$ $v_{11-15}$ $\rightarrow$ $v_{12}$ $\rightarrow$ $v_{13}$,\\

\noindent representing $\{$$c_1$ = 0, $c_2$ = 0, $c_3$ = 0, $c_4$ = 1, $c_5$ = 1, $c_6$ = 0$\}$.\\

$D_4$ can be satisfied by another path:

\vspace{0.2cm}

$v_0$ $\xrightarrow{4,6}$ $v_{11-15}$ $\rightarrow$ $v_{16}$ $\rightarrow$ $v_{17}$,\\

\noindent representing $\{$$c_1$ = 0, $c_2$ = 0, $c_3$ = 0, $c_4$ = 1, $c_5$ = 0, $c_6$ = 1$\}$.\\

The only branching node in this subgraph is $v_{11-15}$ (see Fig. ~\ref{fig3-7}(b)), with respect to which we have three $\textit{RS}$s for $v_0$: $\textit{RS}^{v_{11-15},v_0}_{\{6\}}$[$c_5$] = $\{$$v_{12}$$\}$, $\textit{RS}^{v_{11-15},v_0}_{\{4\}}$[$c_6$] = $\{$$v_{16}$$\}$, and $\textit{RS}^{v_{11-15},v_0}_{\{4,6\}}$[$\$$] = $\{$$v_{13}$, $v_{17}$$\}$. The first two $\textit{RS}$s will be ignored since their sizes are equal to 1.

For the third one, we will create a subgraph as shown in Fig. ~\ref{fig3-7}(d). This is completely identical to the subgraph shown in Fig. ~\ref{fig3-7}(c). From this subgraph, we can get a satisfiable subset $\{$$D_4, D_6$$\}$ satisfied by a path shown below:

\vspace{0.2cm}

$v_0$ $\xrightarrow{4,6}$ $v_{13-17}$,\\

\noindent representing a truth assignment: $\{$$c_1$ = 0, $c_2$ = 0, $c_3$ = 0, $c_4$ = 0, $c_5$ = 0, $c_6$ = 0$\}$.\\

Since the subgraph shown in Fig. ~\ref{fig3-7}(c) is the same as the graph shown in Fig. ~\ref{fig3-7}(d), the recursive call over this subgraph is just a repeated computation, brings no new answer, and should be avoided. In Section III-C, we will discuss how this kind of redundancy can be easily eliminated.\\

The above working process can be summarized as follows.

\vspace{0.2cm}

\begin{itemize}
\item{Search $G$ bottom-up (in postorder).
}

\item{For each branching node, a set of $\textit{RS}$ for different anchor nodes will be created.
}
\item{For each $\textit{RS}$, a trie-like subgraph will be generated. (This can be done by a recursive call of the algorithm).
}
\item{Recursively explore the trie-like subgraph to generate some more subsets of satisfiable conjunctions.



}
\item{For each satisfiable subsets, trace the corresponding path to find its truth assignment.



}
\end{itemize}

In terms of the above discussion, we  come up with a recursive algorithm shown below, in which a data structure $R$ is used to accommodate the result, represented as a set of triplets of the form:

<$\alpha$, $\beta$, $\gamma$>,

\noindent where $\alpha$ stands for a subset of conjunctions, $\beta$ for a truth assignment satisfying the conjunctions in $\alpha$, and $\gamma$ is the size of $\alpha$. Initially, $R$ = $\emptyset$.

\begin{algorithm}
\DontPrintSemicolon
\SetKwInOut{Input}{Input}
\SetKwInOut{Output}{Output}
\Input{a logic formula $C$ in $\textit{CNF}$ with each clause in $C$ containing at most two literals.}
\Output{a largest subset of clauses satisfying a certain truth assignment.}
transform $C$ to another  formula $D$ in $\textit{DNF}$;\\
let $D$ = $D_1$ $\lor$ ... $\lor$ $D_n$;\\
\For{ $i$ = $1$ $\text{to}$ $n$}
{construct a $p$*-graph $P_i^*$ for $D_i$;\\
}
construct a trie-like graph $G$ over $P_1^*$, ..., $P_n^*$;\\

$R$ := $\textit{SEARCH}$($G$);\\
return the result calculated in terms of $R$;

\caption{$\textit{2-MAXSAT}$($C$)}
\end{algorithm}

The input of Algorithm $\textit{2-MAXSAT}$($\;$) is a formula $C$ in $\textit{CNF}$. First, we transform it to another formula $D$ in $\textit{DNF}$ (see line 1). Then, for each $D_i$ in $D$, we will create its $p$*-graph $P_i^*$ (see lines 4). Next, we will contruct a trie-like graph $G$ over all $P_i^*$s (see line 5). In the last step, we call Algorithm $\textit{SEARCH}$($G$) to produce the result (see line 6).

\begin{algorithm}
\DontPrintSemicolon
\SetKwInOut{Input}{Input}
\SetKwInOut{Output}{Output}
\Input{a trie-like subgraphs $G$.}
\Output{a largest subset of conjunctions satisfying a certain truth assignment.}
\If{$G$ is a single $p$*-graph}{$R'$ := subset associated with the leaf node;\\ $R$ := $\textit{merge}$($R$, $R'$);\\return $R$;\\
}
\For{each leaf node $v$ in $G$}{let $R'$ be the  subset associated with $v$;\\ $R$ := $\textit{merge}$($R$, $R'$);}
let $v_1$, $v_2$, ..., $v_k$ be all branching nodes in postorder;\\ 	
\For{ $i$ = $1$ $\text{to}$ $k$}
{let $\textit{RS}_1$, ...,  $\textit{RS}_l$ be all the $\textit{RS}$s constructed for $v_i$;\\
\For{$j$ = $1$ $\text{to}$ $l$}
{
construct a trie-like graph $G'$ in terms of $\textit{RS}_j$;\\
let $u_j$ be the corresponding anchor node;\\
$G''$ := $\{$$u_j$$\}$ $\cup$ $G'$;\\
$R'$ := $\textit{SEARCH}(G'')$;\\ $R$ := $\textit{merge}$($R$, $R'$);
}
}
return $R$;

\caption{$\textit{SEARCH}$($G$)}
\end{algorithm}

Accordingly, the input of Algorithm $\textit{SEARCH}$($\;$) is a  trie-like subgraph $G$. First, we will check whether $G$ is a single $p$*-graph. If it is the case, we must have found a (largest) subset of conjunctions associated with the leaf node, satisfiable by a certain truth assignment (see lines 1 - 4). Since it contains no branching nodes, no further recursive call will be conducted.

Otherwise, we will search $G$ bottom up (in postorder) to find all the branching nodes in $G$. But before that, each subset of conjunctions associated with a leaf node will be first merged into $R$ (see line 5 - 7).

For each branching node $v_i$ encountered, we will first create all its $\textit{RS}$s (see line 10) and for each of them a trie-like  subgraph will be generated (see line 12) and the corresponding anchor node will be added as the virtual node (see line 14). Here, we notice that $G''$ = $\{$$u_j$$\}$ $\cup$ $G'$ is a simplified representation of a complex operation, by which we add not only $u_j$ as a virtual node, but also the corresponding virtual edge and all the inherited edges emitting from $u_j$ to $G'$. Next, a recursive call of the algorithm is made over $G''$ (see linee 15). Finally, the result of the recursive call of the algorithm will be merged into the global answer (see line 16).

Here, the $\textit{merge}$ operation used in line 3, 7, 16 is defined as below.

Let $R$ = $\{$$r_1$, ..., $r_t$$\}$ for some $t$ $\geq$ 0 with each $r_i$ = <$\alpha_i$, $\beta_i$, $\gamma_i$>. We have $\gamma_1$ = $\gamma_2$ = ... = $\gamma_t$. Let $R'$ = $\{$$r_1'$, ..., $r_s'$$\}$ for some $s$ $\geq$ 0 with each $r_i'$ = <$\alpha_i'$, $\beta_i'$, $\gamma_i'$>. We have $\gamma_1'$ = $\gamma_2'$ = ... = $\gamma_s'$. By $\textit{merge}$($R$, $R'$), we will do the following checks.

\begin{itemize}
\item{If $\gamma_1$ < $\gamma_1'$, $R$ := $R'$.}
\item{If  $\gamma_1$ > $\gamma_1'$, $R$ remains unchanged.}
\item{If  $\gamma_1$ = $\gamma_1'$, $R$ := $R$ $\cup$ $R'$.}
\end{itemize}

For simplicity, the above technical details are not described in the algorithm. But it can be easily extended with these operations included.

Besides, to describe the working process of the algorithm, we will introduce a concept of $\textit{RC-trees}$ (or $\textit{recursive-call trees}$), by which the working process of $\textit{REARCH}$($\;$) can be clearly described. In such a $\textit{RC}$-tree, 
we have two kinds of nodes referred to as $\textit{b-nodes}$ and $\textit{RS-nodes}$, respectively. A $b$-node corresponds to a branching node in $T$ while an $\textit{RS}$-node corresponds to a recursive call over an $\textit{RS}$, by which a trie-like subgraph will be constructed.

\vspace{0.2cm}

\noindent $\textbf{Definition}$ $\textbf{4.}$ ($\textit{RC}$-trees) An $\textit{RC}$-tree $\mathcal{T}$ contains two kinds of nodes: $b$-nodes and $\textit{RS}$-nodes. A $b$-node has only $\textit{RS}$-nodes as the children while an $\textit{RS}$-node has only $b$-nodes as the children. We have an edge ($v$, $u$) in $\mathcal{T}$ if $v$ is a  $b$-node and $u$ is an $\textit{RS}$ with respect to a branching node respresented by $v$ (called a $\textit{b-RS}$ edge), or $v$ is an $\textit{RS}$-node and $u$ is a branching node appearing in the trie-like subgraph constructed over all the subgraphs respectively rooted at the nodes in $\textit{RS}$ represented by $v$ (called an $\textit{RS-b}$ edge). The $\textit{root}$ of $\mathcal{T}$ represnts $G$, the initial trie-like graph, which is viewed as a graph created by a $\textit{virtual}$ recursive call over all the initial $p$*-graphs.

\vspace{0.2cm}

As an example, consider the $\textit{RC}$-tree shown in Fig. ~\ref{fig6-1}, by which the whole computation of $\textit{SEARCH}$($G$) over the trie-like graph $G$ (shown in Fig. ~\ref{fig3}) constructed over all the $p$*-graphs for the variable sequences given in Table ~\ref{table1} is described.

\begin{figure*}[ht]
\centering
\includegraphics[width=150mm,scale=1]{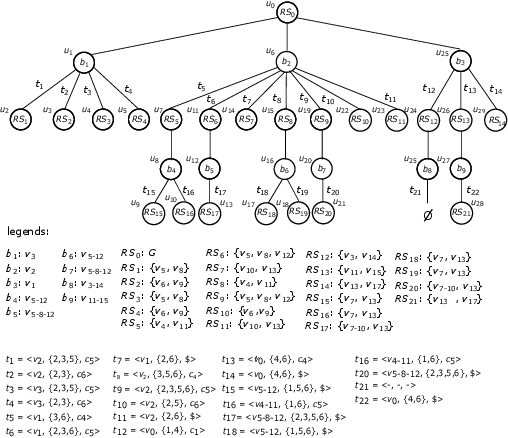}
\caption{An $\textit{RC}$-tree.}
\label{fig6-1}
\end{figure*}

In the $\textit{RC}$-tree $T'$, the $\textit{root}$, $\textit{RS}_{0}$, is a virtual $\textit{RS}$ representing the intial trie-like graph $G$, which contains three branching nodes $v_3$, $v_2$, and $v_1$, represented respectively by the three children of $\textit{RS}_{0}$: $b_1$, $b_2$, and $b_3$ in $\mathcal{T}$. $b_1$ has its own four children $\textit{RS}_1$, $\textit{RS}_2$, $\textit{RS}_3$, and $\textit{RS}_4$, representing four $\textit{RS}$s with repect to $v_3$ in $G$. Since in each of the trie-like subgraphs constructed for these $RS$s there is no branching node (see Example ~\ref{example1}, Fig. ~\ref{fig3-2} and Fig. ~\ref{fig3-3}), each of them has no child. $b_2$ has seven children $\textit{RS}_5$, ... $\textit{RS}_{11}$, corresponding to the seven $\textit{RS}$s with respect to $v_2$ in $G$. Since in the trie-like subgraph constructed for $\textit{RS}_5$ (i.e., $\textit{RS}^{v_2,v_1}_{\{3,6\}}$[$c_4$] = $\{$$v_4$, $v_{11}$$\}$) we have a braching node $v_{5-12}$ (which is the merge of two nodes: $v_5$ and $v_{12}$; see Example ~\ref{example2}, and Fig. ~\ref{fig3-4}), $\textit{RS}_5$ has a child node $b_4$ representing this branching node, which has two children itself, representing two $RS$s with respect to it. Further, in the trie-like subgraph for $\textit{RS}_{6}$ (i.e., $\textit{RS}^{v_2,v_1}_{\{2,3,6\}}$[$c_5$] = $\{$$v_5$, $v_8$, $v_{12}$$\}$), we also have a branching node $v_{5-8-12}$ (which is the merge of three nodes: $v_5$,  $v_8$ and $v_{12}$). Thus, $\textit{RS}_6$ has a child $b_5$, representing $v_{5-8-12}$. Since we have only one $\textit{RS}$ with respect to $v_{5-8-12}$, $b_5$ has only one child $\textit{RS}_{17}$ = $\textit{RS}^{v_{5-8-12},v_{5-8-12}}_{\{1,2,3,6\}}$[$c_1$] = $\{$$v_{7-10}$, $v_{13}$$\}$) (see Fig. ~\ref{fig3-5}). Special attention should be paid to node $\textit{RS}_{12}$ (i.e., $\textit{RS}^{v_1,v_0}_{\{1,4\}}$[$c_1$] = $\{$$v_{3}$, $v_{14}$$\}$). In the trie-like subgraph constructed for this $\textit{RS}$, we have a branching node $v_{3-14}$ (see Fig. ~\ref{fig3-7}(a)). Therefore, it has a child ($b_8$). However, all the $\textit{RS}$s with respect to $v_{3-14}$ are of size equal to 1 and should be ignored. Thus, $b_8$'s child is $\emptyset$.

\vspace{0.2cm}

In the same way, we can check all the other nodes in this $\textit{RC}$-tree.

\vspace{0.2cm}

We can also associate each $\textit{b-RS}$ edge $e$ = ($b$, $\textit{RS}$) with a triplet <$u$, $s$, $c$> to represent a recursive call of the algorithm over the trie-like subgraph constructed in terms of  $\textit{RS}$ (= $\textit{RS}^{v,u}_s$[$c$]), where $v$ is the branching node represented by $b$, and $u$ is an anchor node with respect to $v$, reaching all those nodes with a same label $c$ in different subgraphs in $G$[$v$] (subgraph of $G$ rooted at $v$) through the spans in $s$.

Then, each $\textit{root-to-leaf}$ path in the $\textit{RC}$-tree represents a series  of nested recursive calls shown below, by which a series of trie-like subgraphs $G_1$, ... $G_k$ (for some $k$ $\geq$ 1) will be constructed:

\begin{itemize}
\item[]{<$u_1$, $s_1$, $c_1$> $\rightarrow$ <$u_2$, $s_2$, $c_2$> $\rightarrow$ ... $\rightarrow$ <$u_k$, $s_k$, $c_k$>,
}
\end{itemize}

\noindent where each $u_i$ is an anchor node with respect to a certain branching node in $G_{i-1}$,  $s_i$ is the corresponding label for all the spans connecting $G_{i-1}$ with $G_i$ (represented by the corresponding virtual edge), and $c_i$ for the corresponding variable name. Here, $G_0$ is set to be $G$. We call such a series a chain of recursive calls.


Denote by $w_k$ a leaf node in $G_k$. Assume that $A$ is the subset of conjunctions associated with $w_k$. Then, we need to trace a path $p$ consisting of several subpaths and spans to determine a subset of conjunctions satisfied by a certain truth assignment represented by $p$:

\begin{itemize}
\item[]{$p_0$ $\xrightarrow{s_0}$ $p_1$ $\xrightarrow{s_1}$ $p_2$ ... $\xrightarrow{s_{k-1}}$ $p_k$.
}
\end{itemize}

\noindent where each $p_j$ $\{$$j$ = 0, 1, ..., $k$ - 1$\}$ represents a subpath from the root of $G_{j}$ to the virtual root of $G_{j+1}$ (which is in fact an anchor node in $G_j$), and $s_j$ is the label for the virtual edge in $G_{j+1}$. $p_k$ is the last subpath from the root of $G_k$ to $w_k$. \\

Then, in terms of Lemma ~\ref{lemma3}, the subset of conjunctions satisfied by $p$ is

\vspace{0.2cm}

$A'$ = $A$  $\cap$  $s_1$ $\cap$ ... $\cap$ $s_{k-1}$.

\vspace{0.2cm}

See Fig.~\ref{fig6} for illustration.

\begin{figure}[ht]
\centering
\includegraphics[width=100mm,scale=1]{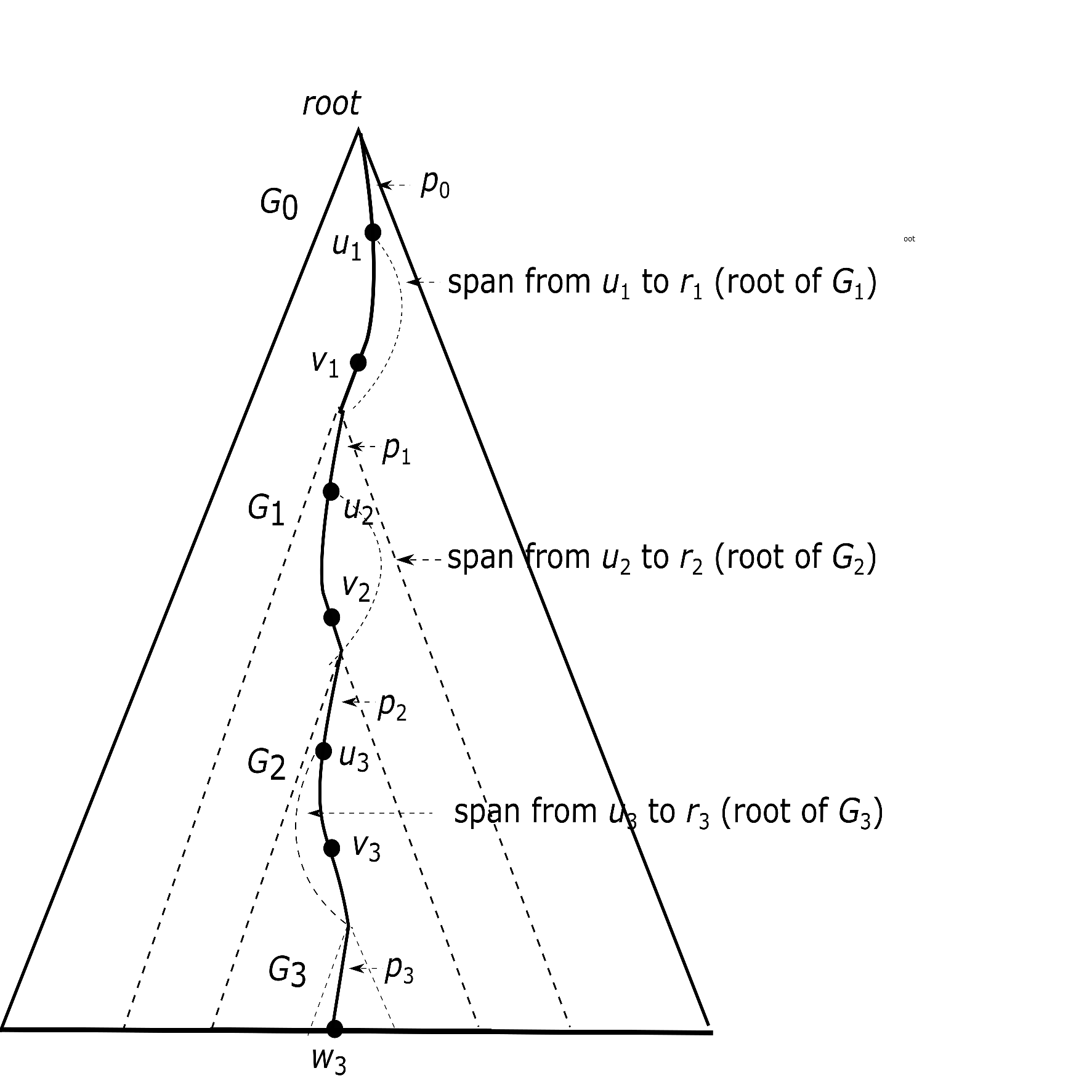} 
\caption{Illustration for tracing truth assignments for satisfied conjunctions.}
\label{fig6}
\end{figure}

In Fig.~\ref{fig6}, we show a chain of three recursivel calls:

\begin{itemize}
\item[]{<$u_1$, $s_1$, $c_1$> $\rightarrow$ <$u_2$, $s_2$, $c_2$> $\rightarrow$  <$u_3$, $s_3$, $c_3$>.
}
\end{itemize}

Here, we assume that $u_1$ is an anchor node with respect to a branching node $v_1$ in $G_0$ (= $G$). By executing <$u_1$, $s_1$, $c_1$>, we will create $G_1$. Further, assume that $u_2$ is an anchor node with respect to a branching node $v_2$ in $G_1$. Then, by executing <$u_2$, $s_2$, $c_2$>, we will generate $G_2$. Next, assume that $u_3$ is an anchor node with respect to a branching node $v_3$ in $G_2$. We will create $G_3$ by executing <$u_3$, $s_3$, $c_3$>. We also assume that $w_3$ is a leaf node in $G_3$, associated with a subset $A$ of conjunctions.

Then, the path shown in Fig.~\ref{fig6} consists of three tree paths $p_i$ (from $\textit{root}_{i}$ to $u_{i+1}$) for $i$ = 0, 1, 2, tree path $p_3$ (from the root of $G_3$ to $w_3$), and three  (combined) spans (from $u_{i}$ to $\textit{root}_{i}$) for $i$ = 1, 2, 3, where $\textit{root}_{i}$ stands for the root of $G_i$.

This path represents a truth assignment satisfying $s$ $\cap$ $A$, where $s$ = $s_1$ $\cap$ $s_2$ $\cap$ $s_3$ and each $s_i$ ($i$ = 1, 2, 3) is the label for the combined span from  $u_{i}$ to $\textit{root}_{i}$.

As an concrete example, consider the branching node $v_2$ in the trie-like graph $G_0$ shown in Fig. ~\ref{fig3} and one of its $\textit{RS}$s: $\textit{RS}^{v_2,v_1}_{\{3,6\}}$[$c_4$] = $\{$$v_4$, $v_{11}$$\}$. In terms of this $\textit{RS}$, $\textit{SEARCH}$($\;$) will create a trie-like subgraph $G_1$ shown in Fig. ~\ref{fig3-4}(a). By a recursive execution of the algorithm on $G_1$, we will generate all the $\textit{RS}$s with respect to its unique branching node $v_{5-12}$. However, for it there is only one $\textit{RS}$: $\textit{RS}^{v_{5-12},v_{4-11}}_{\{1,6\}}$[$\$$] = $\{$$v_7$, $v_{13}$$\}$. In terms of this $\textit{RS}$, the recursive execution of $\textit{SEARCH}$($\;$) will create another trie-like subgraph $G_2$ which contains only a leaf node $v_{17-13}$ (see Fig. ~\ref{fig3-4}(d)) associated with a subset of conjunctions $A$ = $\{$$D_1, D_3, D_5, D_6$$\}$. Adding the corresponding anchor node $v_{4-11}$, we get a graph as shown in Fig. ~\ref{fig3-4}(e). To determine the final subset associated with this leaf node and the corresponding truth assignment, we will trace a path:

\vspace{0.2cm}
$v_0$ $\rightarrow$ $v_{1}$ $\xrightarrow{3,6}$ $v_{4-11}$ $\xrightarrow{1,6}$ $v_{7-13}$,

\vspace{0.2cm}

\noindent which consists of three paths:

\vspace{0.2cm}

- path in $G_0$: $v_0$ $\rightarrow$ $v_1$,

- path in $G_1$:  $v_{4-11}$, (the path contains only a single node)

- path in $G_2$: $v_{7-13}$. (the path contains only a single node)\\

connected through wo spans:

- span $s_1$ connecting $v_1$ and $v_{4-11}$ with label $\{$3, 6$\}$,

- span $s_2$ connecting $v_{4-11}$ and $v_{7-13}$ with label $\{$1, 6$\}$.\\

Then, this path represents a truth assignment $\{$$c_1$ = 0, $c_2$ = 1, $c_3$ = 0, $c_4$ = 1, $c_5$ = 0, $c_6$ = 0$\}$, which satisfies only one conjunction $D_6$ since $A$ $\cap$ $s_1$ $\cap$ $s_2$ = $\{$1, 3, 5, 6$\}$ $\cap$ $\{$3, 6$\}$ $\cap$ $\{$1, 6$\}$ = $\{$6$\}$.








\vspace{0.2cm}

Concerning the correctness of Algorithm $\textit{SEARCH}$($G$), we have the following proposition.

\begin{proposition}
Let $G$ be a trie-like graph established over a logic formula in $\textit{DNF}$. Applying $\textit{SEARCH}$( ) to $G$, we will get a maximum subset of conjunctions satisfying a certain truth assignment.
\label{proposition2}
\end{proposition}
\begin{proof}
To prove the proposition, we first show that any subset of conjunctions found by the algorithm must be satisfied by a same truth assignment. This can be observed by the subsets associated with the leaf nodes of trie-like graphs and subgraphs, as well as the definition of $\textit{RS}$s.

We then need to show that any subset of conjunctions satisfiable by a certain truth assignment can be found by the algorithm. For this purpose,  consider a subset of conjunctions $D'$ = $\{$$D_1$, ..., $D_r$$\}$ ($r$ > 1) which can be satisfied by a truth assignment represented by a path $P$. We will prove by induction on the number $l$ of spans on $P$ that our algorithm is able to find $P$. 

Basic step. When $l$ = 0, $P$ must be a tree path in $T$ and the claim holds. When $l$ = 1, the unique span on $P$ must cover a branching node $v$ in $G$. Let $u$ $\xrightarrow{s}$ $w$ be such a span. Denote by $P'$ the tree path from $\textit{root}$ to $u$ in $T$. Then, by a recursive call of $\textit{SEARCH}$($\;$) to the trie-like subgraph constructed over the corresponding $\textit{RS}^{v,u}_{s'}$[$c$],  we can find a sub-path $P''$; and $P$ must be equal to the concantenation of $P'$, the span $u$ $\xrightarrow{s}$ $v$, and $P''$.

\indent Induction step. Assume that when $l$ = $k$, the algorithm can find $P$.

Now, assume that $P$ contains $k$ + 1 spans $s_1$, $s_2$, ..., $s_k$, $s_{k+1}$. They must corresponds to a chain of $k$ + 1 nested recursive calls of $\textit{SEARCH}$($\;$). Denote by $G_i$
 the trie-like subgraph created by the ($i$ - 1)th recursive call, where $G_0$ = $G$. Let $u$ $\xrightarrow{s}$ $v$ be the first span on $P$. Denote by $P'$ the sub-path from the $\textit{root}$ of $T$ to $u$, and by $P''$ the sub-path of $P$ from $v$ to the last node of $P$. Denote by $D_j$$\backslash$$P'$ the conjunction obtained by removing variables on $P'$ from $D_j$ ($j$ = 1, ..., $r$). Let $D''$ = $\{$ $D_1$$\backslash$$P'$, ...,  $D_r$$\backslash$$P'$$\}$. Then, the truth assignment represented by $P''$ satisfies $D''$. According to the induction hypothesis, $P''$ can be found by executing $\textit{SEARCH}$(\;). Therefore, $P$ can also be found by $\textit{SEARCH}$($\;$). To see this, observe the first recursive call of $\textit{SEARCH}$($\;$) performed when we encounter the first branching node $v$ in $G_0$, by which we will find $P''$ satisfying $D''$. Then, the concantenation of $P'$, the corresponding span (covering $v$), and $P''$ definitely satisfies $D'$. This completes the proof.
\end{proof}

\subsection{Further improvement}\label{subsection3.3}

The algorithm discussed in the previous subsection can be greatly improved. We generally distinguish between two kinds of improvements. The first kind of improvements can be obtained by recognizing idential recursive calls wich may appear along different chains of nested recursive calls. The second kind of improvements can be achieved by imposing some extra controls to suppress any futile recursive calls during a (recursive) execution of $\textit{SEARCH}$($\;$). 

\vspace{0.2cm}
\noindent - $\textit{First kind of improvements}$

\vspace{0.2cm}

In Example 3, we have seen that we may have two recursive calls over a same trie-like subgraph (see the subgraphs respectively shown Fig. ~\ref{fig3-7}(c) and (d) for illustration;  also see identical nodes $\textit{RS}_{14}$ and $\textit{RS}_{21}$ in the $\textit{RS}$-tree $\mathcal{T}$ shown in Fig. ~\ref{fig6-1}). To remove this kind of redundancy, we can use the hashing technique and store the results of a recursive call over a subgraph constructed in terms of an $\textit{RS}_{s}^{v,u}$[$c$] in an address calculated by $\textit{hash}$($\textit{RS}_{s}^{v,u}$[$c$]) in the corresponding global hash table. Thus, to examine whether an input subgraph has been checked before, we need only a constant time to compute a hash value. For instance, before the recursive call for $\textit{RS}^{v_1,v_0}_{\{4,6\}}$[$\$$] = $\{$$v_{13}$, $v_{17}$$\}$ (= $\textit{RS}_{14}$ in $\mathcal{T}$) in Example 3 is carried out, we can first calculate $\textit{hash}$($\textit{RS}^{v_{1},v_0}_{\{4,6\}}$[$\$$]), and find that it is identical to $\textit{hash}$($\textit{RS}^{v_{11-15},v_0}_{\{4,6\}}$[$\$$]) since $\textit{RS}^{v_{11-15},v_0}_{\{4,6\}}$[$\$$] ($\textit{RS}_{21}$ in $\mathcal{T}$) is also equal to $\{$$v_{13}$, $v_{17}$$\}$. Assume that the recursive call for $\textit{RS}^{v_{11-15},v_0}_{\{4,6\}}$[$\$$] has been performed before. Then, the recursive call for $\textit{RS}^{v_1,v_0}_{\{4,6\}}$[$\$$] can be simply ignored.

Due to the redundancy checking, an $\textit{RC}$-tree $\mathcal{T}$ will be reduced to an $\textit{RC}$-graph $\mathcal{G}$, which may contain multiple-edges (two or more edges with both the same end node and the same start node), as illustrated in Fig. ~\ref{fig6-2}. 

\begin{figure*}[ht]
\centering
\includegraphics[width=140mm,scale=1]{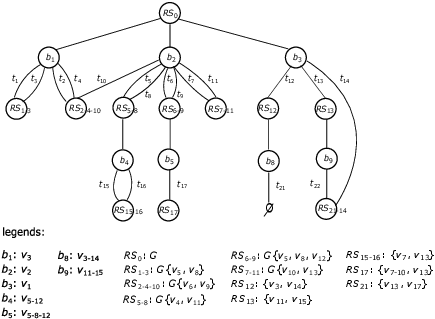}
\caption{An $\textit{RC}$-graph.}
\label{fig6-2}
\end{figure*}

In $\mathcal{G}$ shown in Fig. ~\ref{fig6-2}, each $\textit{b-RS}$ edge is associated with a triple $<$$u$, $s$, $c$$>$ as in $\mathcal{T}$, which enables us to trace the truth assignment satisfying a certain subset of conjunctions, associated with a leaf node in the trie-like subgraph constructed in terms of the corresponging $\textit{RS}$. In addition, a node like $\textit{RS}_{2-4-15}$ shows that $\textit{RS}_{2}$, $\textit{RS}_{4}$ and $\textit{RS}_{15}$ in $\mathcal{T}$ are the same, and therefore only one recursive call needs to be conducted for them. The same analysis applies to all the other nodes in $\mathcal{G}$. Hence, no repeated recursive calls will be performed.

Furthermore, we can also check, each time we encounter a branching node, whether it has been met before. If it is the case, simply skip over it. To see that this type of checking is necessary, too, have a look at Fig. ~\ref{fig10}(a). Denote by $G_1$ the trie-like graph made over the subgraphs respectively rooted at $w_1$ and $w_2$, and by $G_2$ the trie-like graph made over the subgraphs respectively rooted at $v_1$, $v_2$, and  $v_3$. Although $G_1$ and $G_2$ are different, they may contain some common branching nodes. This demonstrates the a same branching node may be encountered multiple times. 

Together with this checking, the check of repeated recursive calls enables us to remove any possible redundant computation.

\begin{figure*}[ht]
\centering
\includegraphics[width=150mm,scale=1]{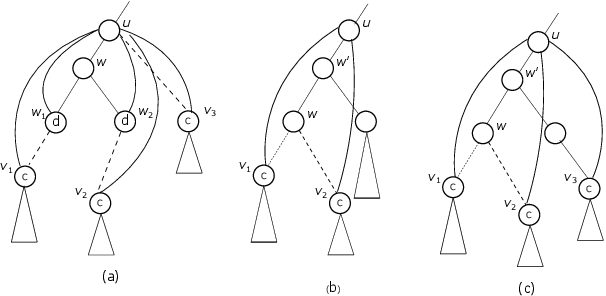} 
\caption{Illustration for  reducing $RS$s and avoiding redundancy.}
\label{fig10}
\end{figure*}


\vspace{0.2cm}
\noindent - $\textit{Second kind of improvements}$

\vspace{0.2cm}

We have different ways to reduce the number of recursive calls.

\begin{itemize}
\item{($\textit{answer size estimation}$) We remember that when generating the trie $T$ over the main paths of the $p$*-graphs created for the variable sequences shown in Table ~\ref{table1}, we have already found a (largest) subset of conjunctions $\{$$D_1$, $D_3$, $D_5$$\}$, which can be satisfied by a truth assignment represented by the corresponding tree path. This is larger than or equal-sized to any found subset of conjunctions which can be satisfied by a certain truth assignment in the subsequent computation. Therefore, all such effort is useless. To avoid this kind of futile work, we can simply perform a pre-checking to see whther a recursive call can possibly generate a larger answer. This can be done by checking the labels for all the virtual edges on the path from $\textit{root}$ to the root of the current trie-like subgraph. For example, before we make a recursive call to create a trie-like subgraph for $\textit{RS}^{v_3,v_2}_{\{2,3,5\}}$[$c_5$] = $\{$$v_5$, $v_8$$\}$ (see Fig. ~\ref{fig3-2}), we can first check whether the size of the span label of this $\textit{RS}$ is smaller than or equal-sized to the largest subset of satisfiable conjunctions already found. Since the size of the span label $\{$2, 3, 5$\}$ is equal to the size of a already found answer the corresponding recursive call should be suppressed.
}
\item(combining $\textit{RS}$s) {All the $\textit{RS}$s with the same label (variable name) can be combined into a single node. We need only to keep the relevant links (edges) to trace the truth assignment and use the associated labels to filter the answers. For example, in Fig. ~\ref{fig6-2}, we can combine $\textit{RS}_{1-3}$ and $\textit{RS}_{6-9}$ into a single node since all the nodes in these two $\textit{RS}$ are labeled with $c_5$. Then, we connec it to two other nodes: $b_1$ using two edges (labeled respectively with $t_1$ and $t_3$), and $b_2$ using two edges (labeled respectively with $t_6$ and $t_9$). These labels enable us to filter conjunctions which cannot be satisfied by the corresponding truth assignment.
}

\item{($\textit{reduction of RSs}$) Consider the graph shown in Fig. ~\ref{fig10}(b). In this graph, we assume that $w$ and $w'$ are two branching nodes in $G$. Then, with respect to $w$ and $w'$, their common ancestor $u$ will have two identical $\textit{RS}$s:\\

$\qquad$ $\textit{RS}_{s}^{w,u}$[C] = $\textit{RS}_{s}^{w',u}$[C] = $\{$$v_1$, $v_{2}$$\}$.\\

Thus, during the execution of $\textit{SEARCH}$($\;$), a same trie-like subgraph will be created two times: one is for $\textit{RS}_{s}^{w,u}$[C] and another is for $\textit{RS}_{s}^{w',u}$[C],  but with the same result to be produced.

$\quad$ However, if we create $\textit{RS}$s only for those anchor nodes appearing on part of a tree path, i.e., on the segment between the current branching node $v$ (including $v$) and its lowest ancestor branching node $v'$ (including $v'$) in $T$, this kind of redudancy can be avoided with possible loss of some answers. But the correctness of the algorithm is not affected since one of the maximum satisfiable subsets of conjunctions can always be found. See Fig. ~\ref{fig10}(c) for illustration. In this figure, the $\textit{RS}$ of $u$ with respect to $w$ is different from the $\textit{RS}$ with respect to $w'$. It is clear that when checking $w$, $\textit{RS}_{s}^{w,u}$[C] will not be computed since $u$ is beyond the segment between $w$ and $w'$. Therefore, the corresponding result will not be generated. But $\textit{RS}_{s}^{w',u}$[C] must cover $\textit{RS}_{s}^{w,u}$[C], implying a larger (or same-sized) subset of conjunctions which can be satisfied by a certain truth assignment.
}
\end{itemize}

\section{Time complexity analysis}

In this section, we analyze the running time of our algorithm for solving a 2-MAXSAT problem with $n$ clauses and $m$ variables.

The total running time consists of three parts.

The first part $\tau_1$ is the time for computing the frenquencies of variable appearances in $D$. Since in this process each variable in a $D_i$ is accessed only once, we have $\tau_1$ = $\mathcal{O}$($NM$), where $N$ = 2$n$ and $M$ = $n$ + $m$. Note that the set of variables for $D$ is $V$ $\cup$ $\{$$x_1$, ..., $x_n$$\}$.

The second part $\tau_2$ is the time for constructing a trie-like graph $G$ for $D$. This part of time can be further partitioned into three portions.

\begin{itemize}
\item{$\tau_{21}$: The time for sorting variable sequences for $D_i$'s. It is obviously bounded by $\mathcal{O}$($NM$log$_2$ $M$).}
\item{$\tau_{22}$: The time for constructing $p$*-graphs for each $D_i$ ($i$ = 1, ..., $N$). Since for each variable sequence a transitive closure over its overlapped spans should be first created and needs $\mathcal{O}$($M^2$) time, this part of cost is bounded by $\mathcal{O}$($NM^2$).}
\item{$\tau_{23}$: The time for merging all $p$*-graphs to form a trie-like graph $G$. This part is also bounded by $\mathcal{O}$($NM^2$).}
\end{itemize}

The third part $\tau_3$ is the time for searching $G$ to find a maximum subset of conjunctions satisfied by a certain truth assignment. It is a recursive procedure. 

First, we show that in all the (recursively) generated trie-like subgraphs, the number of all the branching nodes is bounded by $\mathcal{O}$($NM$).

\begin{lemma}
In all the generated trie-like subgraphs during the excecution of $\textit{SEARCH}$($G$), we have at most $\mathcal{O}$($NM$) different branching nodes.
\label{lemma4}
\end{lemma}

\begin{proof}
From Fig. ~\ref{fig6-2}, we can see that an $\textit{RC}$-graph can be divided into two kinds of levels: $\textit{RS}$-levels and $b$-levels. They are clearly interleaved. Let $h$ be the number of $b$-levels in an $\textit{RC}$-graph. We can see that $h$ $<$ $m$ since the height of a trie-like subgraph at the $i$th $\textit{RS}$-level must be strictly lower than the highest trie-like subgraph at ($i$-1)th $\textit{RS}$-level. Denote by $k_i$ the subset of branching nodes at the $i$th $b$-level. Three of its properties can be observed:

\begin{itemize}
\item{for $i$ $\ne$ $j$, any node in $k_i$ does not appear in $k_j$, and $\textit{vice versa}$.
}
\item{Due to the checking of multiple appearances of an $\textit{RS}$, and multiple apperances of a branching nodes, the nodes in each $k_i$ are different from each other.
}
\item{Each branching node is the merging of a subset of two or more nodes respectively from different $p$*-graphs, which is not overlapped with any other subset for a branching node (due to the combination of $\textit{RS}$s).
}
\end{itemize}

From the first two properties, we know that all the branching nodes generated in the whole working process are different. Then, together with the third property, we show the following inequality:
\[
\sum^h_{\begin{array}{c}\scriptstyle i = 1
\end{array}
}
|k_{i}| \leq NM 
\]
\end{proof}

In the $\textit{RC}$-graph shown in Fig. ~\ref{fig6-2}, we have only 7 branching nodes, much less than $NM$ = 6 $\times$ 6 = 36. (Remember that for the $\textit{DNF}$ formula (2), $N$ = 6 and $M$ = 6.)


With respect to each branching node, we may have at most $M$ anchor nodes on the path from $\textit{root}$ to it; and for each of them, we have at most $M$ $\textit{RS}$s and then $M$ recursive calls. Thus, corresponding to each branching node, $\mathcal{O}$($M^2$) recursive calls of the algorithm may be performed. Since the number of different branching nodes is bounded by $N$$M$, we may have $N$$M$ $\times$ $\mathcal{O}$($M^2$) = $\mathcal{O}$($NM^3$) recursive calls in the worst case.

However, for each branching node $v$, the number of all those anchor nodes for which a trie-like subgraph is actually created, is much less than $\mathcal{O}$($M$). They are just the nodes on a segment between the current branching node $v$ and its lowest ancestor branching node $v'$ in $T$, as discussed in Section III-C (on the second kind of improvements). Denote by $l_v$ the number of the anchor nodes on such a segment (including both $v$ and $v'$). The number of different recursive calls should be bounded by

\[
\sum_{\begin{array}{c}\scriptstyle all\;branching \\[-4pt]
\scriptstyle nodes\; v
\end{array}
}
l_{v} \cdot M = M \cdot 
\sum_{\begin{array}{c}\scriptstyle all\;branching \\[-4pt]
\scriptstyle nodes\; v
\end{array}
}
l_{v} =
\]

\[
M (
\sum_{\begin{array}{c}\scriptstyle all\;branching \\[-4pt]
\scriptstyle nodes\; v
\end{array}
}
(l_{v}-1) +
\sum_{\begin{array}{c}\scriptstyle all\;branching \\[-4pt]
\scriptstyle nodes\; v
\end{array}
}
1)
\leq 2NM^2
\]

It is because corresponding to each node on $l_v$, we have at most $M$ recursive calls.

In terms of the above analysis, we have the following proposition.

\begin{proposition}
Let $G$ be a trie-like graph over a formula in $\textit{DNF}$ containing $N$ conjunctions with $M$ variables. The time complexity of Algorithm $\textit{SEARCH}$($G$) is bounded by $\mathcal{O}$($N^2M^4$).
\label{proposition5}
\end{proposition}
\begin{proof}
In terms of Lemma ~\ref{lemma4}, in the whole working process we will check at most $\mathcal{O}$($NM$) branching nodes in the worst case, and altogether make $\mathcal{O}$($NM^2$) recursive calls. By each recursive call, $\mathcal{O}$($NM^2$) time is required to generate a trie-like subgraph. So, the time complexity of $\textit{SEARCH}$($G$) is bounded by $\mathcal{O}$($NM^2$) $\times$ $\mathcal{O}$($NM^2$) = $\mathcal{O}$($N^2M^4$).
\end{proof}

Since $N$ = 2$n$ and $M$ = $n$ + $m$, we have $\mathcal{O}$($N^2M^4$) = $\mathcal{O}$($n^2(n+m)^4$).

\section{Conclusions}
In this paper, we have presented a new method to solve the 2-MAXSAT problem. The worst-case time complexity of the algorithm is bounded by  $\mathcal{O}$($n^2(n+m)^4$), where $n$ and $m$ are respectively the numbers of clauses and variables of a logic formula $C$ (over a set $V$ of variables) in $\textit{CNF}$ with each clause containing at most 2 literals. The main idea behind this is to construct a different formula $D$ (over a different set $U$ of variables) in $\textit{DNF}$, according to $C$, with the property that for a given integer $n$* $\leq$ $n$ $C$ has at least $n$* clauses satisfied by a truth assignment for $V$ if and only if $D$ has least $n$* conjunctions satisfied by a truth assignment for $U$. To find a truth assignment that maximizes the number of satisfied conjunctions in $D$, a graph structure, called a $p$*-graph, is introduced to represent each conjunction in $D$. In this way, all the conjunctions in $D$ can be represented as a trie-like graph $G$. Searching $G$ bottom up in a recursive way, we can find the answer efficiently. Since 2-MAXSAT is $\textit{NP}$-complete, our algorithm in fact provides a proof of $P$ = $\textit{NP}$.\\

\vspace{1pt}
\begin{IEEEbiography}[{\includegraphics[width=1in,height=1.25in,clip,keepaspectratio]{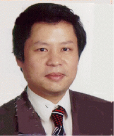}}]{Dr. Yangjun Chen}got his PhD in Computer Science from the University of Kaiserslautern, Germany, in 1995. He is now a professor in Dept. Applied Com-puter Science, the University of Winnipeg, Canada. He has published more than 200 scientific articles  in Computer Science and Computer Software engineering.
\end{IEEEbiography}

\end{document}